\documentclass[twocolumn]{aastex631}

\usepackage{CJK}
\usepackage{verbatim} % FOR BLOCK COMMENTS
\usepackage[T1]{fontenc}
\usepackage[utf8]{inputenc}
\usepackage{graphicx}
\usepackage{xcolor}
\usepackage{amsmath}
\usepackage{amssymb}
\usepackage{booktabs}
\usepackage{enumitem}
\usepackage{microtype}
\usepackage{tikz}

\newcommand{\msun}{\ensuremath{M_{\odot}}}

\usepackage{xspace}
\newcommand{\sfr}{\ensuremath{\mathrm{SFR}}}
\providecommand{\arcsec}{^{\prime\prime}}
\providecommand{\arcmin}{^{\prime}}
\providecommand{\fs}{.\!^{\mathrm{s}}}
\providecommand{\farcs}{.\!^{\prime\prime}}

\begin{document}
\begin{CJK*}{UTF8}{gkai}
%\title{Constraining cosmological variation of fundamental constants towards PKS 1830-211 }
\title{ASKAP discovery of a pair of large radio bubbles: on the origin of odd radio circles}

\author[0000-0002-2432-2587]{Renzhi Su (苏仁智)}
\email{rzsu.astro@gmail.com}
\affiliation{Shanghai Astronomical Observatory, Chinese Academy of Sciences, 80 Nandan Road, Shanghai 200030, China}

\author[0000-0002-4455-6946]{Minfeng Gu (顾敏峰)}
\email{gumf@shao.ac.cn}
\affiliation{Shanghai Astronomical Observatory, Chinese Academy of Sciences, 80 Nandan Road, Shanghai 200030, China}

\author[0000-0001-8416-7059]{Hengxiao Guo}
\affiliation{Shanghai Astronomical Observatory, Chinese Academy of Sciences, 80 Nandan Road, Shanghai 200030, China}

\author[0009-0005-7553-049X]{Neelesh Amrutha}
\affiliation{Research School of Astronomy and Astrophysics (RSAA), Australian National University, Canberra ACT 2611, Australia}

\author[0000-0002-1908-0536]{Liang Chen}
\affiliation{Shanghai Astronomical Observatory, Chinese Academy of Sciences, 80 Nandan Road, Shanghai 200030, China}

\author[0000-0003-3922-5007]{Fangzheng Shi}
\affiliation{INAF - Osservatorio Astronomico di Roma Via Frascati, 33 I-00078 Monte Porzio Catone (RM), Italy}

\author[0000-0002-4521-6281]{Wenwen Zuo}
\affiliation{Shanghai Astronomical Observatory, Chinese Academy of Sciences, 80 Nandan Road, Shanghai 200030, China}

\author[0000-0002-6563-7438]{Ruiyu Zhang}
\affiliation{School of Physics, Henan Normal University, Xinxiang 453000, China}

\author[0000-0002-0092-7944]{Guobin Mou}
\affiliation{Department of Physics and Institute of Theoretical Physics, Nanjing Normal University, Nanjing 210023, People’s Republic of China}

\author[0000-0003-1474-8899]{Fulai Guo}
\affiliation{Shanghai Astronomical Observatory, Chinese Academy of Sciences, 80 Nandan Road, Shanghai 200030, China}

\author[0000-0002-3742-6609]{Wenke Ren}
\affiliation{Shanghai Astronomical Observatory, Chinese Academy of Sciences, 80 Nandan Road, Shanghai 200030, China}

\author[0000-0002-1564-0436]{Xuechen Zheng}
\affiliation{Shanghai Astronomical Observatory, Chinese Academy of Sciences, 80 Nandan Road, Shanghai 200030, China}



\begin{abstract}
We report the serendipitous discovery of a pair of large, low-surface-brightness radio bubbles in
944\,MHz continuum data from the ASKAP Evolutionary Map
of the Universe (EMU) survey. The structure, centred on the elliptical galaxy LEDA~217397 at a
redshift of $z=0.040$, spans $\sim$399~kpc and comprises two partly overlapping shells with radii of
$\sim$114~kpc, connecting common kpc-scale radio bubbles to odd radio circles (ORCs). Its integrated
flux density is $56.8\pm2.9$\,mJy, implying a
rest-frame 1.4\,GHz luminosity of $1.4\times10^{23}$\,W\,Hz$^{-1}$. Combining the EMU data with
MWA GLEAM-X at 88--185\,MHz, we derive a steep spectrum
($\alpha=-1.04\pm0.04$) and a spectral-index map suggesting possible exterior flattening. SED
fitting indicates a massive ($\log M_{\ast}/\msun = 10.97\pm0.09$), quiescent
($\sfr=0.025\pm0.083\,\msun$\,yr$^{-1}$) early-type host with no mid-infrared AGN signature and
no overdense environment. We compare the bubbles with ORCs and large radio
shells, and discuss three scenarios for their origin: a starburst-driven wind, a merger-driven
shock, and AGN jet-inflated bubbles. The starburst wind is disfavoured on energetic grounds
($\gtrsim$$10^{59}$\,erg required versus $\sim$$10^{8}$\,yr electron lifetimes), and neither a
halo-scale merger shock nor a spherical blast wave can explain the double-shell geometry; a
bipolar nuclear outburst -- a relic AGN jet episode, perhaps triggered by a
supermassive-black-hole merger -- provides the most natural explanation, with later shocks possibly
re-energising the plasma. Deeper broad-band radio, polarimetric, spectroscopic and X-ray
observations are needed to confirm its nature.

\end{abstract}



\section{Introduction} \label{sec:intro}

Wide-field radio continuum surveys carried out with a new generation of sensitive telescopes have revealed a population of faint, circular, edge-brightened radio sources that do not fit any previously known class of object. The first examples, dubbed ``odd radio circles'' (ORCs), were found in the Evolutionary Map of the Universe (EMU) Pilot Survey conducted with the Australian Square Kilometre Array Pathfinder (ASKAP) \citep{Norris2021a}, with one further case identified in archival Giant Metrewave Radio Telescope (GMRT) data and another, ORC~J0102$-$2450, in additional ASKAP observations \citep{Koribalski2021}. ORCs typically appear as rings or edge-brightened discs roughly one arcminute across; where a central galaxy can be identified, the rings correspond to shells several hundred kiloparsecs in diameter \citep{Norris2021a,Norris2021b,Rupke2024}. Confirmed and candidate ORCs have since been reported in ASKAP, MeerKAT, GMRT, VLA and LOFAR data
\citep{Gupta2022,Lochner2023,Kumari2024a,Kumari2024b,Bulbul2024,Koribalski2024a,
Koribalski2024b,Gupta2025,Norris2025,Hota2025,Taziaux2025,Filipovic2026}.

Deep follow-up observations have established the characteristic properties of the class. MeerKAT imaging of ORC1 resolved a thin, edge-brightened ring with a well-ordered, tangential magnetic field and a steep radio spectrum \citep[$\alpha\simeq-1.4$ between 88 and 1284\,MHz;][]{Norris2022}. The central galaxies are massive ($M_{\ast}\sim10^{11}\,\msun$), red, unobscured ellipticals with old stellar populations \citep{Rupke2024}, and at least one of them is surrounded by an extended, kinematically disturbed [O\,{\sc ii}] nebula $\sim$40\,kpc across, hinting at a recent energetic event in or near the host \citep{Coil2024,Coil2026}. The known ORCs tend to reside in over-dense environments or to have close companions, which may indicate that the ambient medium plays a role in lighting up the shells \citep{Norris2021b}.

A wide range of formation scenarios have been proposed, most of them involving a spherical or bipolar outflow seen at a favourable orientation. These include a blast wave from a cataclysmic event in the host galaxy, such as the merger of two supermassive black holes \citep[SMBHs;][]{Koribalski2021,Norris2022} or the cumulative effect of tidal disruption events \citep{Omar2022b}; the termination shock of a powerful starburst-driven wind or its fossil remnant \citep{Norris2021a,Norris2022,Fujita2024}; merger-driven shocks travelling through the circumgalactic medium (CGM) of galaxies and groups \citep{Dolag2023,Bulbul2024,Koribalski2024a,Koribalski2024b,Koribalski2026a,Ivleva2026}; virial shocks around massive halos \citep{Yamasaki2024}; remnant lobes of radio galaxies re-energised by shocks \citep{Shabala2024}; cosmic-ray dominated, jet-inflated AGN bubbles seen end-on \citep{Lin2024}; supernova remnants in the intragroup medium of the Local Group and its neighbours \citep{Omar2022a}; and even Galactic foregrounds, which are largely ruled out by the high Galactic latitudes and extragalactic redshifts \citep{Norris2021a}. Related phenomena, such as bipolar outflows from nearby disc galaxies \citep{Koribalski2026b} and circularly symmetric emission around ellipticals \citep{Kumari2024a}, suggest that large radio shells may be a more general by-product of galaxy evolution, of which ORCs are an extreme manifestation.

In this paper we present the discovery of a pair of large radio bubbles centred on the elliptical galaxy LEDA~217397 at a redshift of $z=0.040$. At a projected size of $\sim$399\,kpc it falls squarely in the size range of ORCs, while its two-component shell morphology connects it to the growing family of ORC-like radio shells around nearby galaxies \citep{Koribalski2024b,Koribalski2026a}. We first identified the bubbles in the continuum image of the First Large Absorption Survey in HI \citep[FLASH;][]{Allison2022}, an HI spectral survey carried out with ASKAP, which can also provide continuum images. However, the bubbles were observed in EMU as well. We then used the public EMU data for all the analyses since that provides higher sensitivity and spatial resolution. Section~\ref{sec:obs} describes the ASKAP and MWA data; Section~\ref{sec:results} presents the radio properties and spectral behaviour of the bubbles; Section~\ref{sec:discussion} discusses the host galaxy and the possible physical origins; Section~\ref{sec:summary} summarises our findings.

Throughout, we
assume a flat $\Lambda$ cosmology with $\Omega_{\mathrm{m}}=0.3$ and
$H_{0}=70$\,km\,s$^{-1}$\,Mpc$^{-1}$.

\section{Observations and data reduction}\label{sec:obs}

\subsection{ASKAP EMU}\label{sec:obs_flash}

The radio data of this work come from the EMU \citep[][]{Norris2011}, an ASKAP survey whose primary goal is to obtain a deep, wide-field continuum census of the radio sky. ASKAP, located at the Murchison Radio-astronomy Observatory in Western Australia, consists of 36 antennas of 12\,m diameter, each equipped with a
phased-array feed that forms 36 dual-polarisation beams and delivers an instantaneous
field of view of $\sim$30\,deg$^{2}$ \citep{Hotan2021}.

The field containing our target was observed on 15 May 2025 as part of the EMU
programme (scheduling block SB73778). The data were processed by the
ASKAP science data pipeline \textsc{ASKAPsoft} \citep{Guzman2019,wieringa2020}, which performs
bandpass, gain and flagging calibration for each of the 36 formed beams, followed by
deconvolution, self-calibration, primary-beam correction and linear mosaicking; the
calibrated visibilities, continuum images and cubes are served through the CSIRO ASKAP
Science Data Archive \citep[CASDA;][]{Huynh2020}. For the analysis presented here we used
the continuum image centred at 944\,MHz, which has been convolved to a circular
restoring beam of $15\arcsec\times15\arcsec$ with an rms (Root Mean Square) noise of 30\,$\mu$Jy\,beam$^{-1}$
in the vicinity of the target.

\subsection{MWA GLEAM-X}\label{sec:obs_mwa}

Low-frequency coverage is provided by the GaLactic and Extragalactic All-sky Murchison
Widefield Array survey eXtended \citep[GLEAM-X;][]{HurleyWalker2022}, conducted with the
Murchison Widefield Array \citep[MWA;][]{Tingay2013}. GLEAM-X covers the sky south of
Declination $+30^{\circ}$ in the 72--231\,MHz range with a resolution of
$\sim$45$^{\prime\prime}$. Observations are taken as 2-minute snapshots using the drift-scan observing mode, snapshots are then calibrated and mosaicked together to create deeper images of the entire Southern sky\footnote{Further details on the processing steps are described in the GLEAM-X survey description paper  \citep{HurleyWalker2022} and the processing pipeline is available on GitHub (https://github.com/GLEAM-X/GLEAM-X-pipeline)}. GLEAM-X data are produced
in five 30.72\,MHz-wide sub-bands centred at 88, 118, 154, 185 and 216\,MHz. The formally
released data regions (GLEAM-X DR1 and DR2; \citealt{HurleyWalker2022,Ross2024}) do not
cover our target, so we requested the data from the GLEAM-X team, which will be part of upcoming GLEAM-DR4 covering the remaining extra-galactic sky (Ross et al. in prep.). However, due to the low Declination of our source and its location near the South Celestial Pole, it was not covered by the highest GLEAM-X band centred at 216 MHz. For the spectral-index analysis, the EMU image and the 154\,MHz GLEAM-X image
were both convolved to a common circular beam of $90\arcsec\times90\arcsec$ and regridded
to the same pixel scale.

\subsection{ANU optical spectroscopy}\label{sec:ANU_opt}
An optical spectrum of LEDA 217397 was obtained on 28 July 2026 using the Wide Field Spectrograph \citep[WiFeS;][]{dopita_hart_2007,dopita_rhee_2010} on the ANU 2.3-m telescope at Siding Spring Observatory. WiFeS is an integral field unit (IFU) with a field of view of $38\times25$ arcsec$^2$ sampled by $1$ arcsec$^2$ spaxels. Observations consisted of two 700 s exposures using the B3000 and R3000 gratings, providing wavelength coverage of 3200--9800\AA\ at a spectral resolution of $R\simeq 3000$. The data were reduced with an updated version of the \texttt{PyWiFeS} pipeline \citep{childress_vogt_2014_pywifes}, which performs standard CCD reduction including bias subtraction, flat-fielding, wavelength calibration using arc-lamp exposures, telluric correction and flux calibration using spectrophotometric standard stars, and cube reconstruction. A spectrum was extracted from the IFU cube with a 6 arcsec diameter aperture after subtraction of the local sky background. From the spectrum, we measured a redshift of 0.040.

\subsection{Ancillary optical and infrared data}\label{sec:obs_ancillary}

To identify counterparts and study the host galaxy we used $g$, $r$, $i$ and $z$-band
imaging from the DESI Legacy Imaging Surveys Data Release 10 \citep{Dey2019}, $J$, $H$
and $K_{\mathrm{s}}$-band imaging from the Two Micron All Sky Survey Extended Source
Catalogue \citep[2MASS XSC;][]{Skrutskie2006,Jarrett2000} and $W1$, $W2$, $W3$
mid-infrared imaging from the Wide-field Infrared Survey Explorer
\citep[WISE; WISE J094937.14-855304.6;][]{Wright2010,Cutri2013}.
Photometric redshifts of galaxies in the wider field were taken from the neural-network
photometric-redshift catalogue of \citet{Tian2026}, based on DESI Legacy Imaging Surveys
and Pan-STARRS data.

\subsection{Host-galaxy photometry}\label{sec:obs_phot}

To construct the spectral energy distribution (SED) of LEDA~217397 we performed our own
aperture photometry on the images described above. For each band we downloaded the
calibrated image cutout centred on the galaxy and measured the flux within a circular
aperture of radius 28\,arcsec, which comfortably encloses the galaxy's optical extent.
Nearby contaminating sources were masked
out before the flux was summed, and the local background was estimated from a
surrounding source-free annulus. The measured fluxes were converted to AB magnitudes, and all magnitudes were finally
corrected for foreground Galactic extinction using the dust maps of
\citet{Schlegel1998}. The resulting photometry in ten bands ($g$, $r$, $i$, $z$, $J$,
$H$, $K_{\mathrm{s}}$, $W1$, $W2$, $W3$) is listed in Table~\ref{tab:phot} and was used
for the SED fitting in Sect.~\ref{sec:host}.

\section{Results}\label{sec:results}

\subsection{Detection of a pair of radio bubbles}\label{sec:detection}

\begin{figure*}[t]
\gridline{\fig{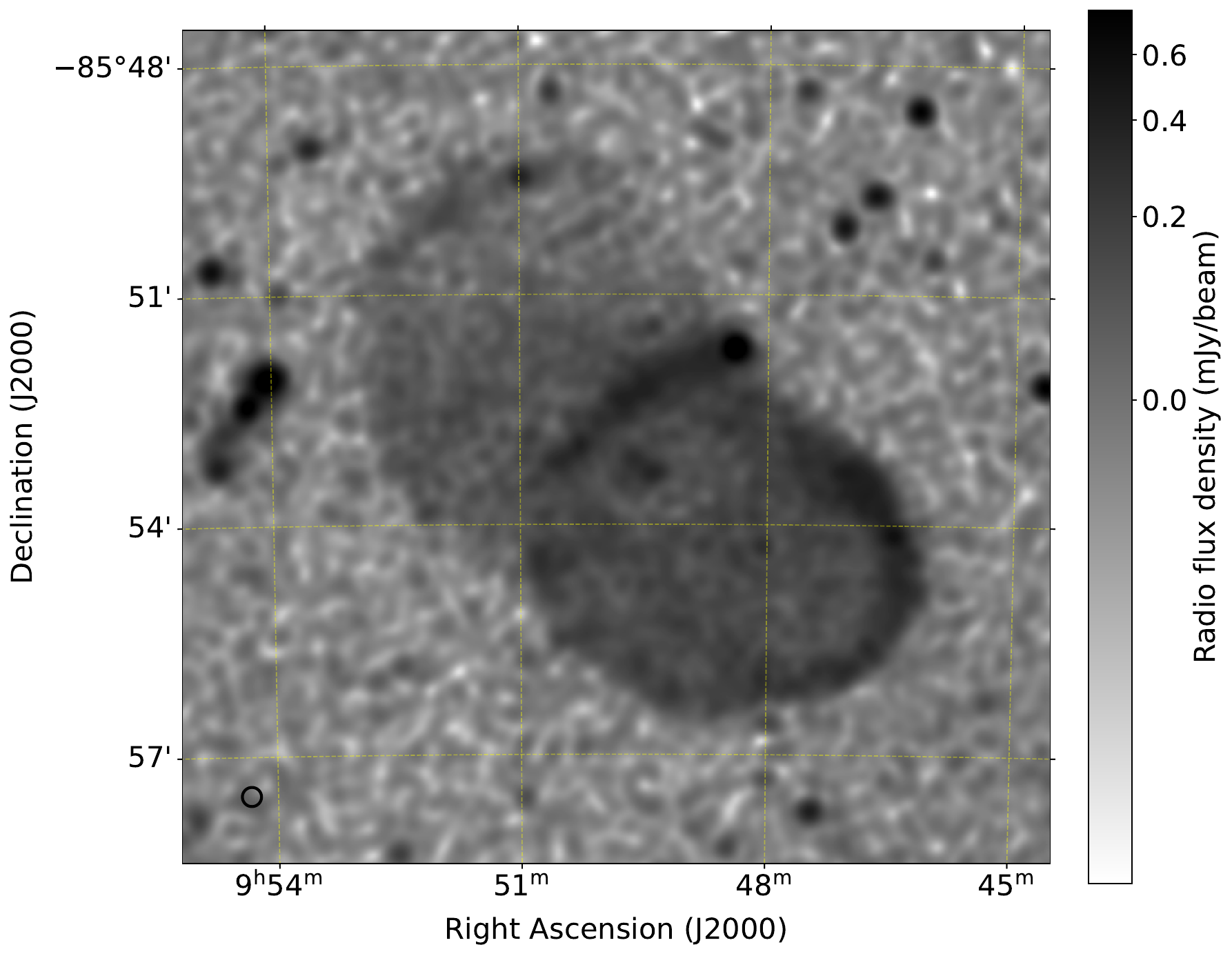}{0.4\textwidth}{(a)}
          \fig{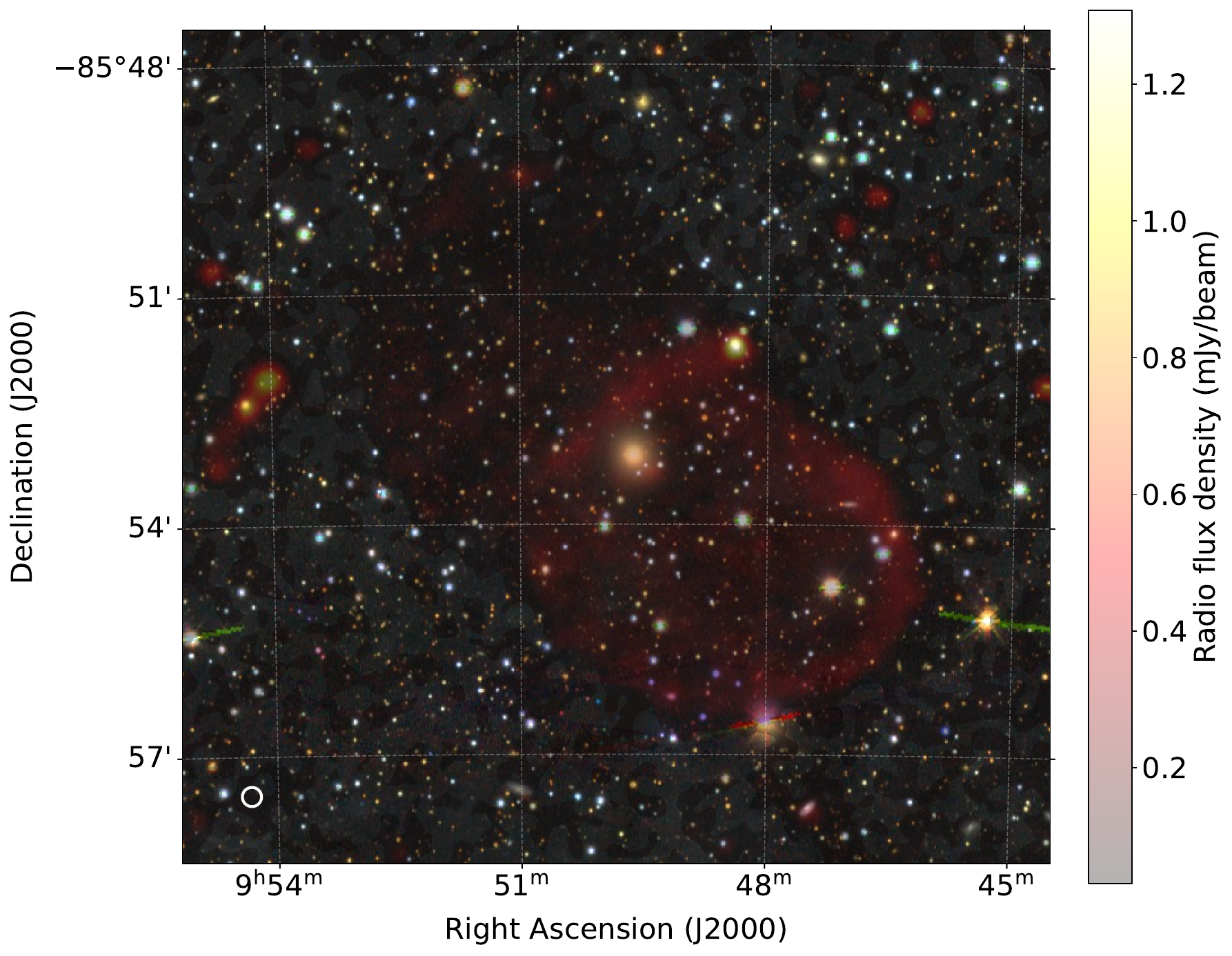}{0.4\textwidth}{(b)}}
\gridline{\fig{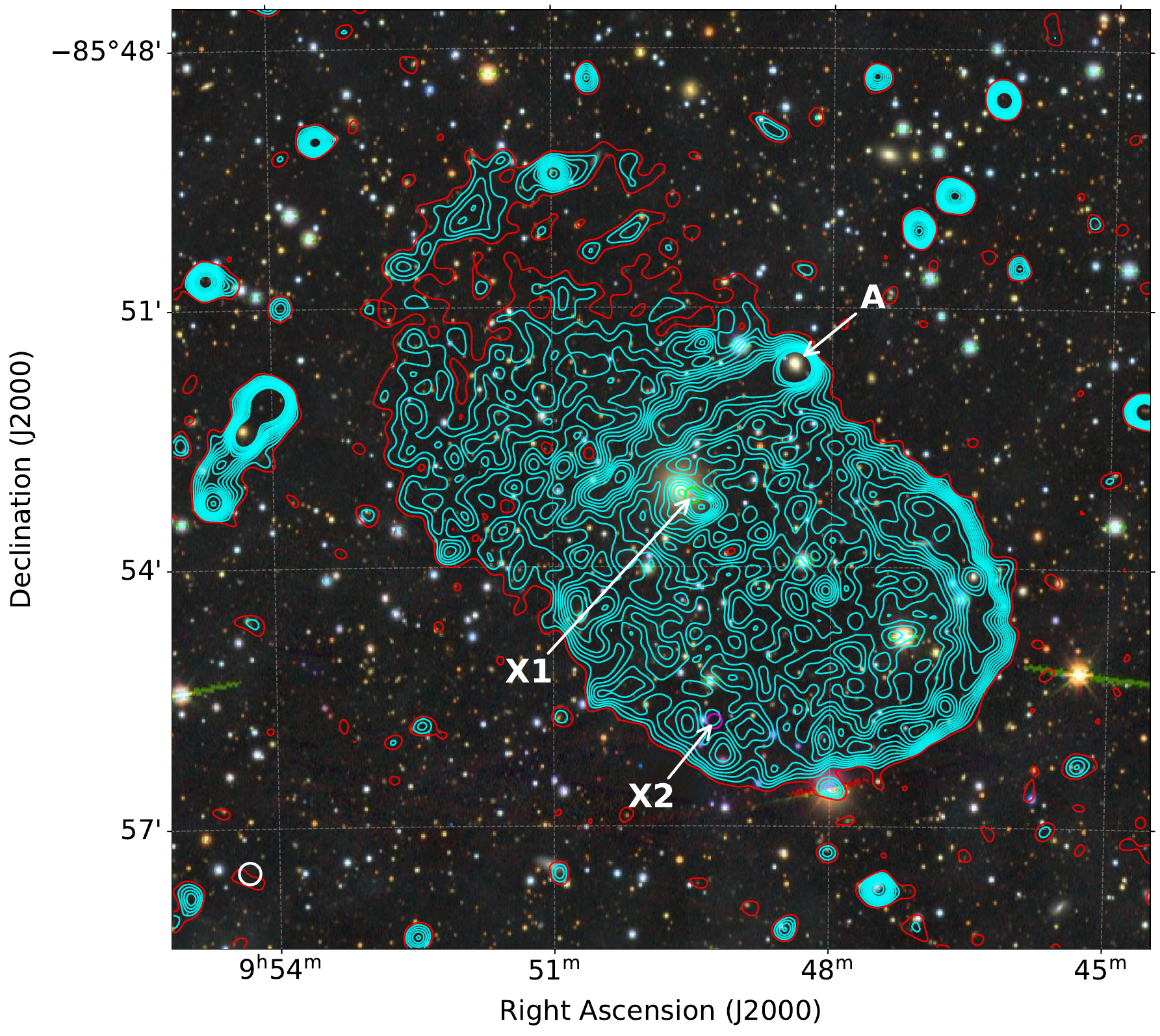}{0.6\textwidth}{(c)}}
    \caption{The radio bubbles in the EMU SB73778 field. (a) ASKAP EMU
    944\,MHz image, convolved to a $15\arcsec\times15\arcsec$ circular beam (shown in
    the lower-left corner); the local rms noise is 30\,$\mu$Jy\,beam$^{-1}$. (b) Colour
    overlay of the EMU emission (red--yellow) on the DESI Legacy Imaging Surveys
    optical image. (c) Contour overlay of the EMU image on the same optical image;
    contour levels are 1 (red), 2, 3, 4, 5, 6, 7, 8, 9, 10, 12, 14 and 16 times the rms. The host galaxy LEDA~217397 lies at the centre of the whole structure;
    the background galaxy ``A'' ($z\simeq0.08$) is marked. Two eROSITA soft X-ray sources X1 and X2 within the bubbles are marked with green and purple circles of which radii represent position uncertainties. }
    \label{fig:bubble}
\end{figure*}

We identified a striking, large-scale diffuse
radio structure centred close to the elliptical galaxy LEDA~217397
(RA~$=09^{\mathrm{h}}49^{\mathrm{m}}37\fs1489$,
Dec~$=-85^{\circ}53\arcmin04\farcs628$, J2000). Figure~\ref{fig:bubble} presents three
views of the source: the 944\,MHz EMU image itself, a colour overlay of the radio
emission on the DESI Legacy Imaging Surveys optical image, and a contour overlay of the
same data. The structure spans $\sim$8.4\,arcmin at its largest extent and shows a
clumpy, limb-brightened shell morphology,
while the interior is filled with fainter emission several times the noise level.

LEDA~217397, with a spectroscopic redshift of $z=0.040$, sits almost at the
geometrical centre of the whole radio structure. Within this global symmetry the galaxy
is offset from the centres of the two shells described below (see Section \ref{sec:size_and_mor}), by
$\simeq$105\,arcsec from the SW shell centre and
$\simeq$112\,arcsec from the NE shell centre, and lies close to the line joining
the two shell centres. This combination, a host at the
centre of the overall radio structure but between its two sub-components, makes LEDA~217397 the
natural host candidate. We emphasise that the probability of a chance alignment between a
bright elliptical galaxy and the centre of such a coherent structure is small, analogous
to the argument made for the original ORCs \citep{Norris2021a,Norris2021b}. The galaxy
labelled ``A'' in Fig.~\ref{fig:bubble}(c), located to the north-west of the host and
partially blended with the bright radio ridge, has a photometric redshift of $z\simeq0.08$
and is therefore a background object; it contributes only $\simeq1.9$\,mJy ($\sim$3\,per
cent) to the integrated flux density of the bubbles and does not affect any of our
conclusions. The association between galaxy “A” and the radio ridge remains unclear. Although they may be physically connected, as in the case of one-sided jets \citep[e.g.][]{norris2025b}, they could also be unrelated, in which case the radio ridge would be part of a brightened ring energized by shocks. Higher spatial resolution and multi-band radio observations will help clarify their relationship. We note that the radio ridge has a flux density of $\sim$ 5 mJy, which will not significantly affect our subsequent analysis regardless of whether it is included. 

No other optical or infrared counterpart to the diffuse radio emission is
found, mirroring the defining property of ORCs that the shells are invisible at non-radio
wavelengths \citep{Norris2021a,Norris2022}.

\subsection{Radio properties}\label{sec:properties}

\subsubsection{Size and morphology}\label{sec:size_and_mor}

\begin{figure*}[t]
\gridline{\fig{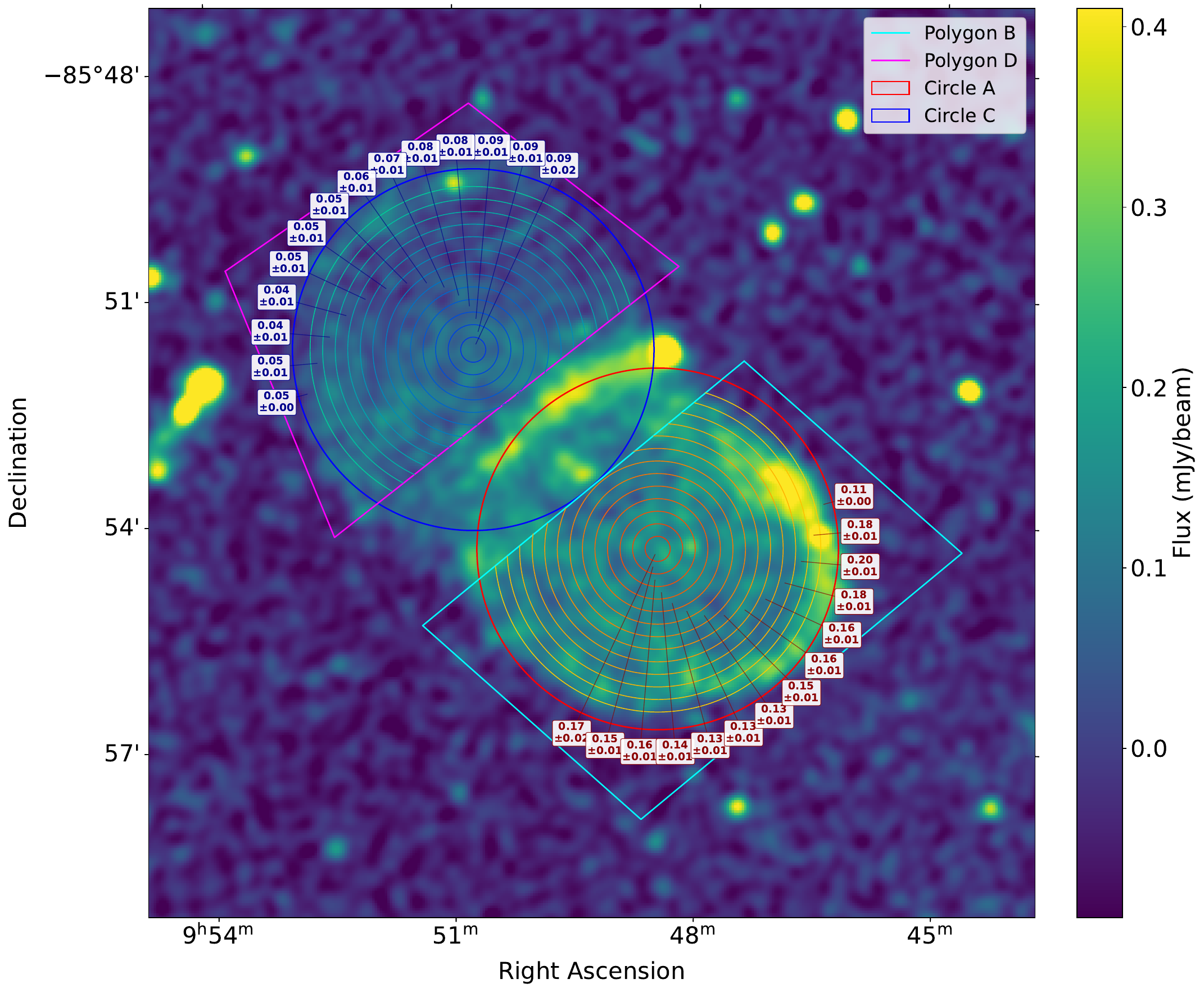}{0.9\textwidth}{(a)}}
\gridline{\fig{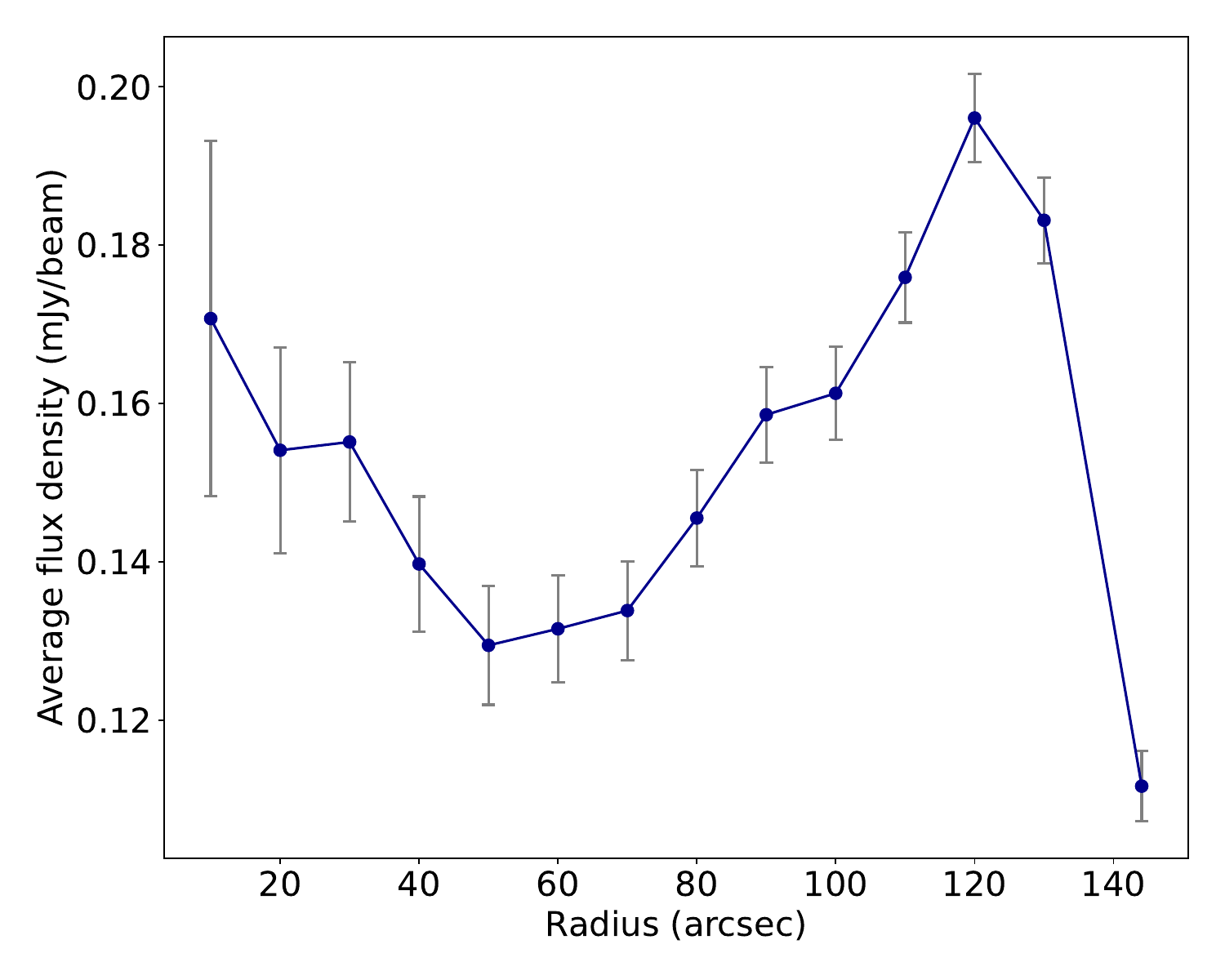}{0.45\textwidth}{(b)}
          \fig{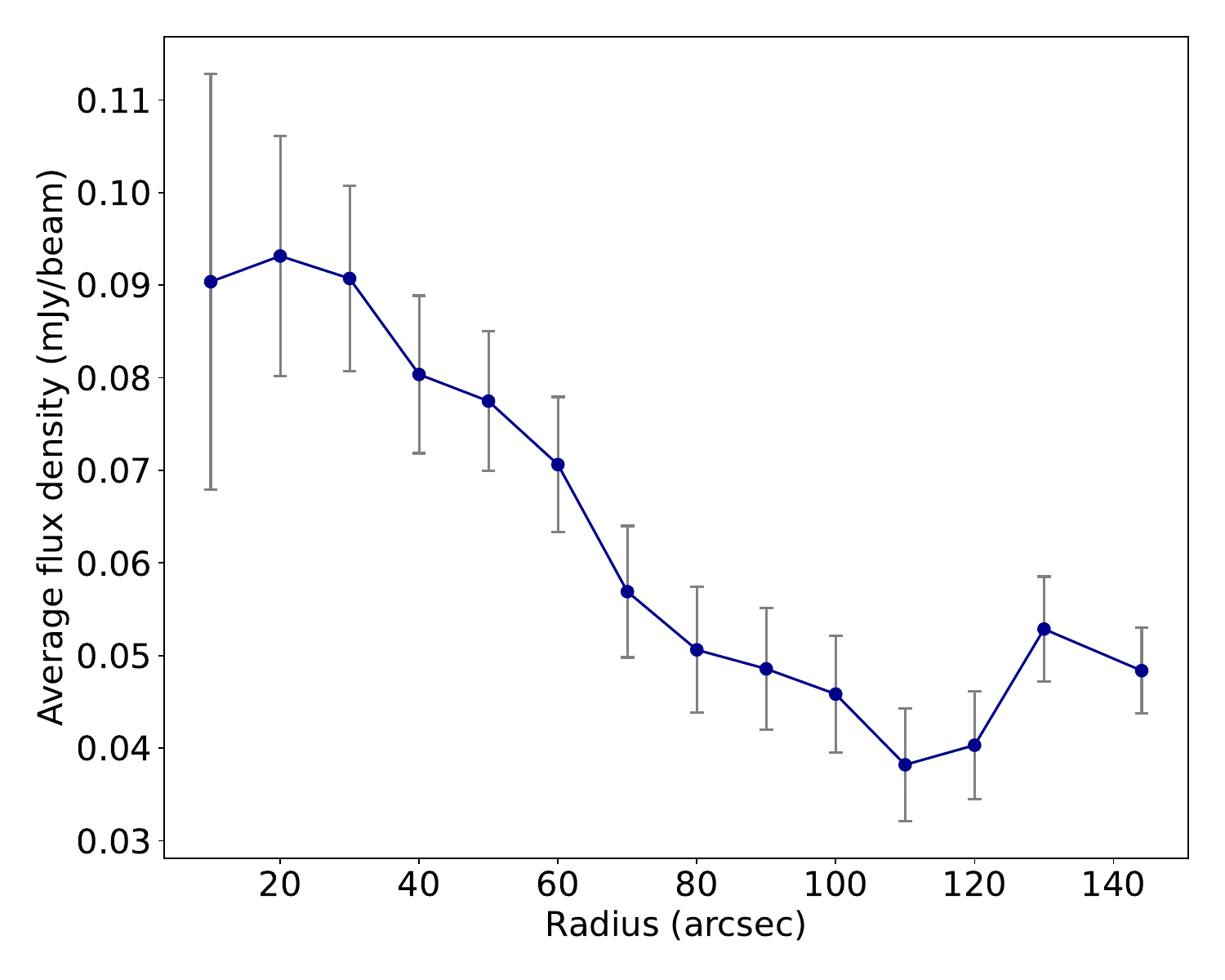}{0.45\textwidth}{(c)}}
    \caption{Shell geometry and radial surface-brightness profiles of the bubbles. (a) The radio morphology could be well described as two partly overlapping circles A and C. The radio emission within Polygon B and D and corresponding concentric annuli was used to derive the radial surface-brightness profiles. The mean flux density (in mJy beam$^{-1}$) within annuli is marked aside.
    (b) Mean 944\,MHz surface brightness in those annuli as a function of radius for the SW component;
    error bars show $\sigma/\sqrt{N_{\rm beams}}$. (c) Same as (b) for the NE component.}
    \label{fig:profiles}
\end{figure*}

The radio structure is not a single symmetric ring, but we found that it could be well described as two partly
overlapping, circular shells. The south-western (SW) circle has a centre at RA~$=09^{\mathrm{h}}48^{\mathrm{m}}28\fs64$, Dec~$=-85^{\circ}54\arcmin19\farcs39$ and a radius of $144^{\prime\prime}$, i.e.\ a diameter of $\sim$228\,kpc at $z=0.040$; the north-eastern (NE) circle has a centre at
RA~$=09^{\mathrm{h}}50^{\mathrm{m}}45\fs27$, Dec~$=-85^{\circ}51\arcmin40\farcs8$ with
the same radius of $144^{\prime\prime}$ ($\sim$228\,kpc diameter), see Fig.~\ref{fig:profiles}a. The largest extent of the combined
structure is $\sim$8.4\,arcmin, corresponding to $\sim$399\,kpc. The elongation of the
structure runs roughly along the line joining the two shell centres (SW--NE).

\subsubsection{Flux density, surface brightness and luminosity}

Using a polygon aperture drawn around the source on the EMU
image, we measure a total integrated flux density of $S_{944}=56.8\pm2.9$\,mJy where the uncertainty takes into account the rms noise and the flux calibration uncertainty of 5\%; the background
galaxy ``A'' contributes only $\simeq1.9$\,mJy and is included in, but negligible for,
this measurement. The mean surface brightness of the bubbles is $\simeq$1.2\,mJy\,arcmin$^{-2}$ ($\simeq$0.35\,$\mu$Jy\,arcsec$^{-2}$), comparable to the
surface brightness of the known ORCs \citep[$\sim$50--200\,$\mu$Jy\,beam$^{-1}$ in ASKAP
images;][]{Norris2021a}. At a luminosity distance of 176.5\,Mpc, the measured flux
density corresponds to a monochromatic luminosity of
$L_{944}=(2.1\pm0.1)\times10^{23}$\,W\,Hz$^{-1}$; extrapolating to 1.4\,GHz with the
measured spectral index (Sect.~\ref{sec:spix}) gives $S_{1.4}\simeq38$\,mJy and $L_{1.4}\simeq1.4\times10^{23}$\,W\,Hz$^{-1}$. Given the small redshift, it can be directly taken as rest-frame luminosity. 

\subsubsection{Brightness distribution: limb brightening}

The averaged surface-brightness profiles are shown in Fig.~\ref{fig:profiles}b, c. For the SW component, the mean surface brightness declines from
$\simeq$0.15--0.17\,mJy\,beam$^{-1}$ in the innermost bins to a minimum of
$\simeq$0.13\,mJy\,beam$^{-1}$ at radii of $\sim$50--70\,arcsec, then rises to a
peak of $\simeq$0.20\,mJy\,beam$^{-1}$ at $\sim$120\,arcsec. The ring is
therefore brighter than the interior by $\sim$50\,per cent at the resolution of the EMU image; given the $15\arcsec$ beam, the true limb-to-interior contrast is likely
higher. The NE component is fainter (declining from $\simeq$0.09\,mJy\,beam$^{-1}$ in the
centre to $\simeq$0.04--0.05\,mJy\,beam$^{-1}$ near the edge), indicating a more filled morphology. The
edge-brightened character matches the defining
morphological signature of ORCs \citep{Norris2021a,Norris2022}.

\subsubsection{Comparison with ORCs and large radio shells}

At $\sim$400\,kpc across, the radio structure sits comfortably within the 300--500\,kpc diameter
range established for the original ORCs \citep{Rupke2024,Norris2021b}, and it is larger
than the $\sim$145\,kpc shells of the nearby Physalis system \citep{Koribalski2024b} and
the $\sim$25\,kpc horseshoe ring of J1407+0453 \citep{Kumari2024b}, while remaining well
below the $\sim$1\,Mpc extent of the most extreme ORC-like structures
\citep{Hota2025,Dolag2023}. Its 944\,MHz luminosity of $2.1\times10^{23}$\,W\,Hz$^{-1}$
is at the lower end of, but consistent with, the radio powers of several
$10^{23}$\,W\,Hz$^{-1}$ inferred for the classical ORC population
\citep{Dolag2023,Norris2022}. In morphology, the double-shell appearance is reminiscent
of the two intersecting shells interpreted as galaxy-merger shock relics by
\citet{Koribalski2026a} and of the paired shells around the interacting Physalis galaxy
pair \citep{Koribalski2024b}. With $z=0.040$, the structure is, after the Physalis system
($z=0.017$), one of the nearest ORC-like objects known, and at $S_{944}\simeq57$\,mJy
it is by far the brightest in terms of observed flux density (the classical ORCs have
$S_{944}\sim2$--10\,mJy; \citealt{Norris2021a,Koribalski2021,Norris2025}), making it a uniquely
accessible target for detailed follow-up. The basic properties of the source are
summarised in Table~\ref{tab:props}.

\begin{table}[t]
\caption{Basic properties of the radio bubbles and its host.}
\label{tab:props}
\centering\small
\begin{tabular}{ll}
\toprule
Parameter & Value \\
\midrule
Host galaxy & LEDA~217397 \\
Host position (J2000) & $09^{\rm h}49^{\rm m}37\fs15$, $-85^{\circ}53\arcmin04\farcs6$ \\
Spectroscopic redshift & 0.040 \\
Luminosity distance & 176.5\,Mpc \\
Angular scale & 0.79\,kpc\,arcsec$^{-1}$ \\
Largest angular size & $\sim$8.4\,arcmin ($\sim$399\,kpc) \\
SW shell radius & $144^{\prime\prime}$ ($\simeq$114\,kpc) \\
NE shell radius & $144^{\prime\prime}$ ($\simeq$114\,kpc) \\
$S_{944}$ (total) & $56.8\pm2.9$\,mJy \\
$L_{944}$ & $(2.1\pm0.1)\times10^{23}$\,W\,Hz$^{-1}$ \\
$L_{1.4\,{\rm GHz}}$ & $\simeq1.4\times10^{23}$\,W\,Hz$^{-1}$ \\
Spectral index $\alpha$ & $-1.04\pm0.04$ \\
Host $\log(M_{\ast}/\msun)$ & $10.97\pm0.09$ \\
Host SFR & $0.025\pm0.083\,\msun$\,yr$^{-1}$ \\
\bottomrule
\end{tabular}
\begin{minipage}{\columnwidth}\vspace{2pt}\footnotesize
Note: $L_{1.4\,{\rm GHz}}$ is the observer-frame 1.4\,GHz luminosity extrapolated from
944\,MHz with the measured $\alpha$. Given the small redshift, it can be directly taken as rest-frame luminosity. 
\end{minipage}
\end{table}

\begin{table}[t]
\caption{Aperture photometry of the host galaxy LEDA~217397 (28\,arcsec radius aperture; see Sect.~\ref{sec:obs_phot}). Magnitudes are in the AB system and are corrected for Galactic extinction following \citet{Schlegel1998}.}
\label{tab:phot}
\centering\small
\begin{tabular}{llc}
\toprule
Survey & Band  & $m_{\rm AB}$ (mag) \\
\midrule
DESI DR10 & $g$ & $15.24\pm0.07$ \\
DESI DR10 & $r$  & $14.41\pm0.05$ \\
DESI DR10 & $i$  & $14.04\pm0.04$ \\
DESI DR10 & $z$& $13.78\pm0.04$ \\
2MASS  & $J$ &  $13.42\pm0.06$ \\
2MASS  & $H$  & $13.19\pm0.07$ \\
2MASS  & $K_{\rm s}$ & $13.48\pm0.08$ \\
WISE & $W1$  & $14.16\pm0.04$ \\
WISE & $W2$  & $14.81\pm0.04$ \\
WISE & $W3$ & $15.63\pm0.09$ \\
\bottomrule
\end{tabular}
\end{table}

\subsection{Spectral index}\label{sec:spix}

\subsubsection{Integrated spectrum}

\begin{figure}[t]
    \centering
    \includegraphics[width=0.95\columnwidth]{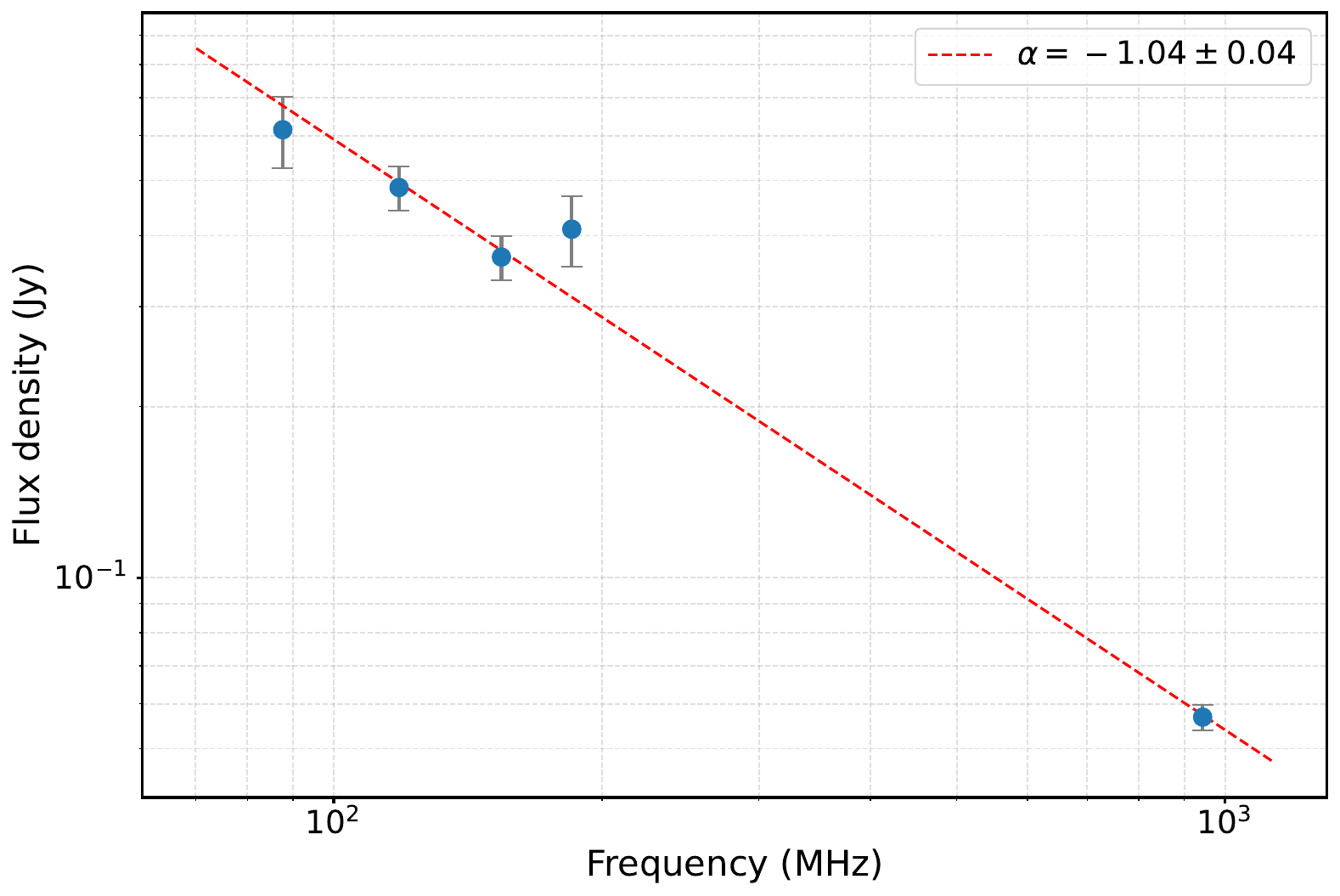}
    \caption{Integrated radio spectrum of the bubbles. Blue points: polygon-flux
    measurements from the four GLEAM-X wide-band images (88--185\,MHz) and the EMU
    944\,MHz image. Dashed red line: best-fitting single power law with $\alpha=-1.04\pm0.04$.}
    \label{fig:spec}
\end{figure}

The bubbles are clearly detected in all four GLEAM-X wide-band images, with polygon-flux
measurements of $0.66\pm0.08$, $0.55\pm0.04$, $0.44\pm0.03$ and $0.44\pm0.06$\,Jy at 88,
118, 154 and 185\,MHz, respectively. The background galaxy ``A'' is very weak in the
GLEAM-X images and contributes negligibly at these frequencies. Combining the four
GLEAM-X points with the EMU 944\,MHz flux density and fitting a single power law
$S_{\nu}\propto\nu^{\alpha}$ yields $\alpha=-1.04\pm0.04$, see Fig.~\ref{fig:spec}. The spectrum is thus steep and, within the
errors, shows no strong curvature in frequency.

This integrated spectral index is similar to those measured for ORC1, both from its
multi-band spectrum between 88 and 1284\,MHz ($\alpha=-1.4\pm0.05$;
\citealt{Norris2022}) and from the original EMU Pilot analysis of its diffuse emission
($\alpha=-1.17\pm0.04$; \citealt{Norris2021a}), and it is considerably steeper than the
$\alpha\simeq-0.5$ expected for freshly accelerated electrons at shocks. Such
steep spectra are a common feature of the ORC population and are generally interpreted as
evidence for an aged or re-accelerated relativistic electron population
\citep{Norris2022,Shabala2024,Dolag2023}.

\subsubsection{Spatially resolved spectral index}

\begin{figure*}[t]
    \centering
    \includegraphics[width=0.9\textwidth]{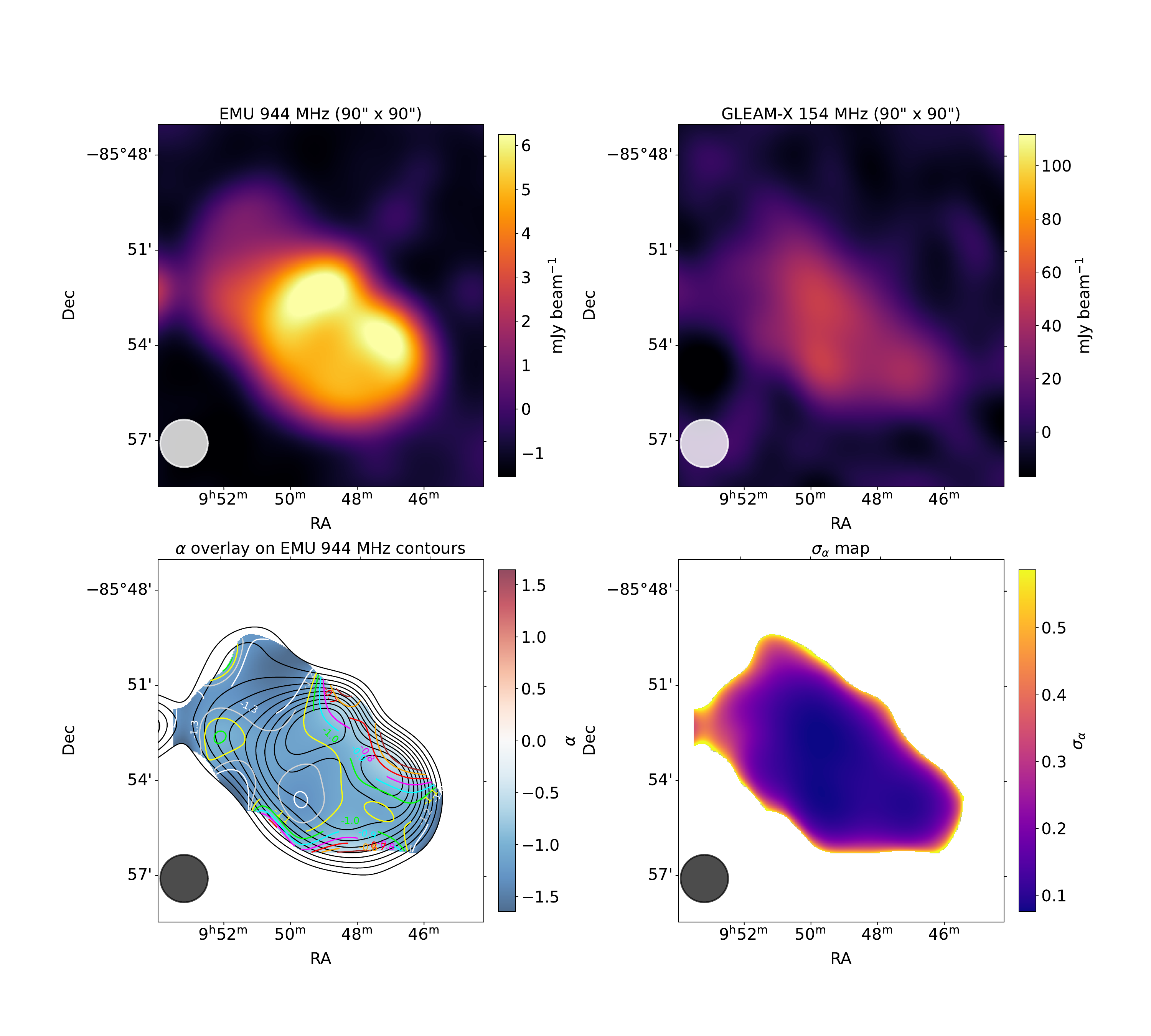}
    \caption{Spectral-index analysis. Top left: EMU 944\,MHz image convolved to
    $90\arcsec\times90\arcsec$. Top right: GLEAM-X 154\,MHz image at the same
    resolution. Bottom left: spectral-index map $\alpha_{154}^{944}$ (colour) overlaid
    on EMU contours, using only pixels above $1\sigma$ in both images. Bottom right:
    corresponding $1\sigma$ uncertainty map; errors grow rapidly towards the periphery.}
    \label{fig:alpha}
\end{figure*}

To search for spectral structure across the bubbles we convolved the EMU image and the
GLEAM-X 154\,MHz image to a common $90\arcsec\times90\arcsec$ resolution and constructed
a pixel-by-pixel spectral-index map between 154 and 944\,MHz, retaining only pixels
detected above $1\sigma$ in both images (Fig.~\ref{fig:alpha}). The median spectral index
over the $\sim$3.1$\times10^{4}$ valid pixels is $-1.13$, consistent with the integrated value, with a median uncertainty of 0.15 per pixel.

A notable overall trend is that the interior spectral index is steeper than the exterior spectral index. It can be as steep as -1.3 in the interior, flattening down to $\sim-0.5$--0.6 at the edge. This is consistent with a scenario in which electrons are (re-)accelerated or compressed near
the shell boundary. The hint of re-steepening at part of the edge could reflect synchrotron and inverse-Compton ageing of electrons diffused ahead of the
shock, or a decline in acceleration efficiency as the shell decelerates \citep{drury1983,Ivleva2026}. However, the uncertainties of the external spectral indices are large, due to the sensitivity limitation. Deeper multi-band observations at matched-resolution are required to confirm and exploit the
spatially resolved spectrum of the bubbles.

\section{Discussion}\label{sec:discussion}

\subsection{The host galaxy}\label{sec:host}

The putative host, LEDA~217397, is a bright elliptical galaxy at $z=0.040$. DESI
Legacy Surveys photometry ($g=15.24$, $r=14.41$, $i=14.04$, $z=13.78$; AB;
Table~\ref{tab:phot}) yields colours of $g-r=0.83$, $r-i=0.37$ and $i-z=0.26$. Such red
optical colours are characteristic of the red sequence occupied by quiescent, early-type
galaxies with old stellar populations \citep[e.g.][]{Bell2003,Strateva2001}. The WISE
mid-infrared colours are $W1-W2=-0.65$ and $W2-W3=-0.82$ (AB, corresponding to
$\simeq0.0$ and $\simeq1.0$ in the Vega system): the former is far below
the $W1-W2\geq0.8$ (Vega) criterion for radiatively efficient AGN \citep{Stern2012}, and both
colours are typical of passive ellipticals whose mid-infrared
emission is dominated by the old stellar population
\citep{Jarrett2011}. There is thus no infrared evidence for a currently active black hole
or for dust-enshrouded star formation in the host.

\begin{figure*}[t]
    \centering
    \begin{tikzpicture}
        \node[anchor=south west, inner sep=0] (img) {\includegraphics[width=0.9\textwidth]{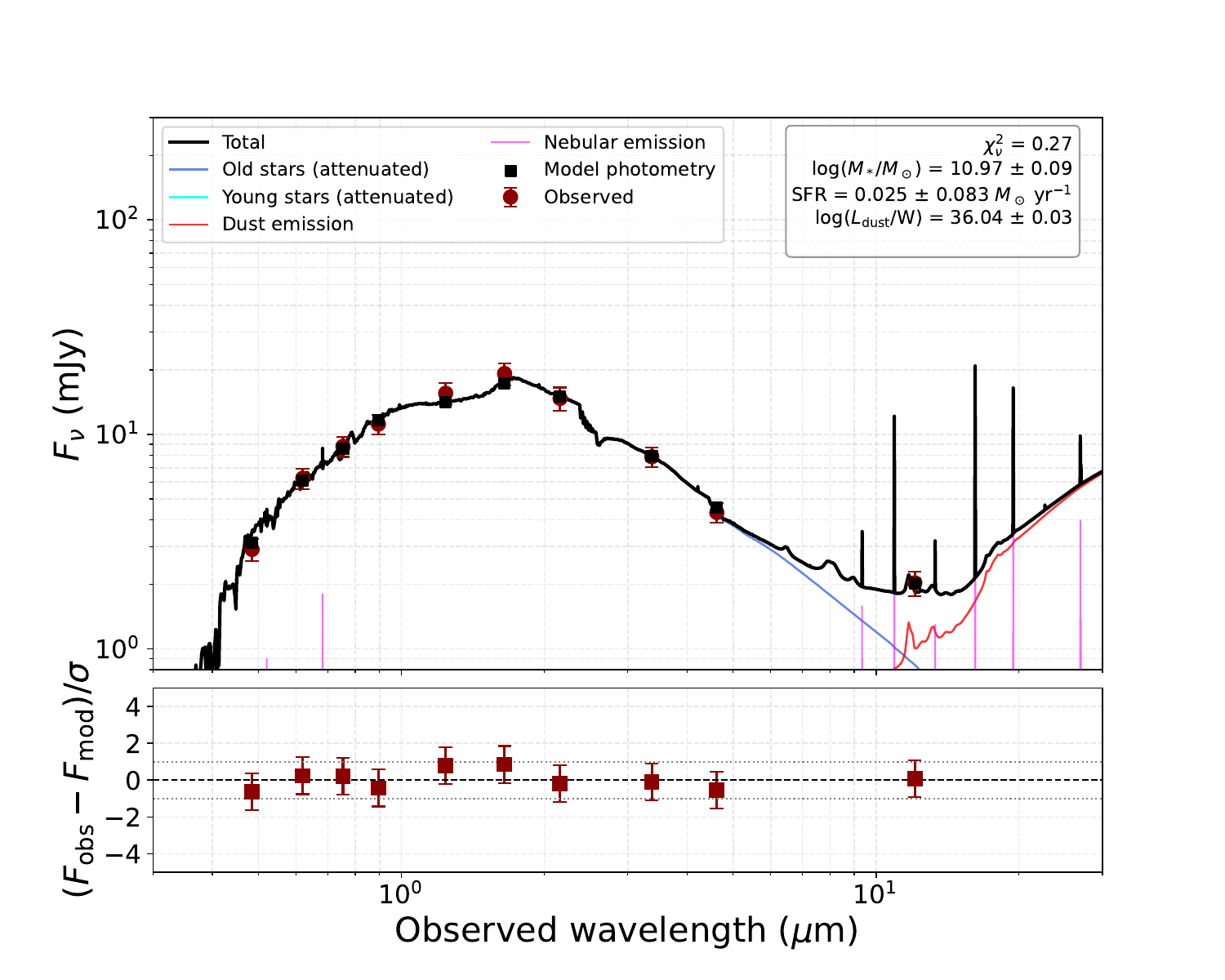}};
    \end{tikzpicture}
    \caption{\textsc{cigale} SED fit to the DESI Legacy Surveys $griz$, 2MASS
    $JHK_{\mathrm{s}}$ and WISE $W1W2W3$ photometry of LEDA~217397. The best fit gives
    $\log(M_{\ast}/\msun)=10.97\pm0.09$, $\sfr=0.025\pm0.083\,\msun$\,yr$^{-1}$ and
    $\log(L_{\mathrm{dust}}/\mathrm{W})=36.04\pm0.03$, with reduced $\chi^{2}=0.27$.}
    \label{fig:sed}
\end{figure*}

\begin{figure}[t]
    \centering
    \includegraphics[width=0.9\columnwidth]{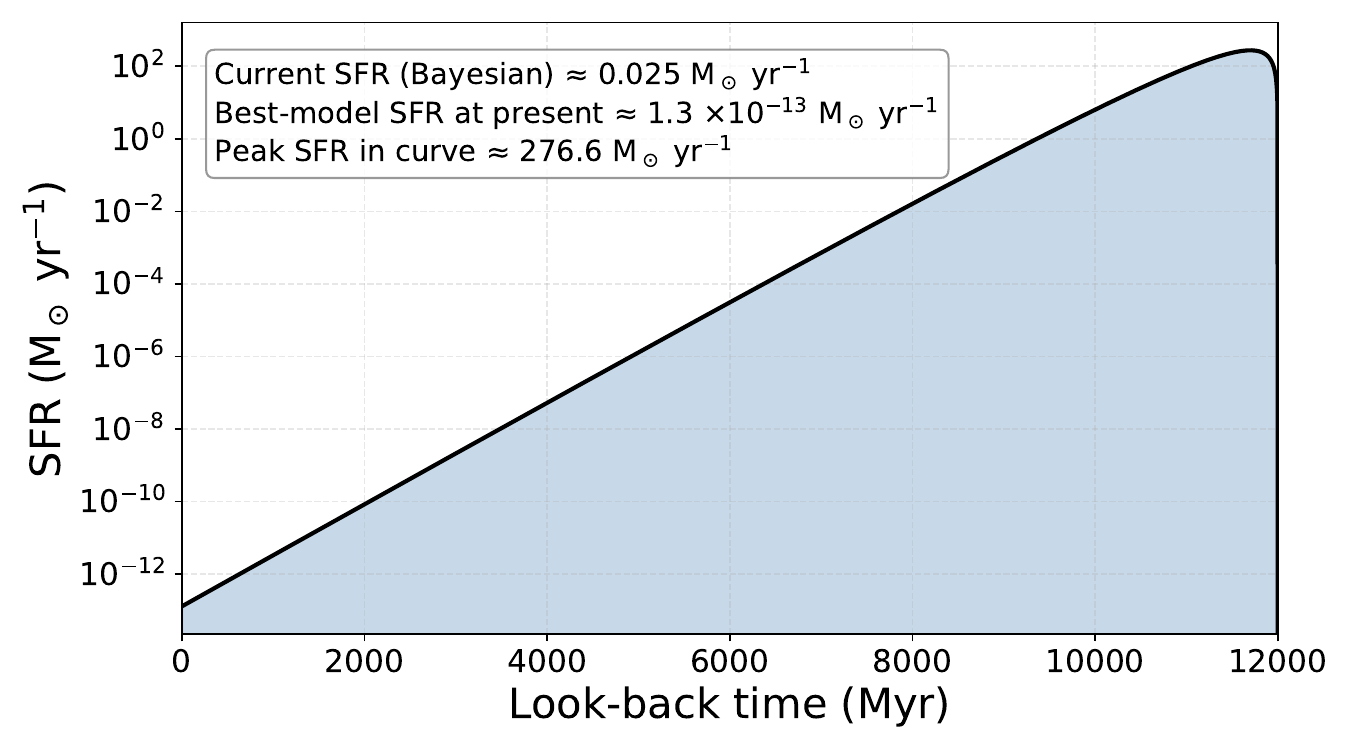}
    \caption{The star formation history of LEDA 217397.}
    \label{fig:sfh}
\end{figure}

We fitted the $grizJHK_{\mathrm{s}}+W1W2W3$ spectral energy distribution (SED;
Sect.~\ref{sec:obs_phot}, Table~\ref{tab:phot}) with
\textsc{cigale} \citep{Boquien2019}, assuming a delayed
exponential star-formation history. The results are shown in Fig.~\ref{fig:sed} and Fig.~\ref{fig:sfh}.

The best fit (reduced $\chi^{2}=0.27$) gives a stellar mass of $\log(M_{\ast}/\msun)=10.97\pm0.09$, a
present-day star-formation rate of $\sfr=0.025\pm0.083\,\msun$\,yr$^{-1}$ and a dust
luminosity of $\log(L_{\mathrm{dust}}/\mathrm{W})=36.04\pm0.03$. The specific star-formation
rate, $\simeq$$3\times10^{-13}$\,yr$^{-1}$, is far below
the star-forming main sequence at this redshift (adopting the main-sequence
calibration of \citealt{Speagle2014}, which gives ${\rm sSFR}\sim10^{-11}$\,yr$^{-1}$ at
this stellar mass and redshift), and the best-fit star-formation history (Fig.~\ref{fig:sfh})
rises towards early cosmic times, indicating that the bulk of the stars formed more than
$\sim$10\,Gyr ago. LEDA~217397 is therefore a massive and quenched elliptical galaxy.

These properties are strikingly similar to those of the confirmed ORC hosts, which are
likewise massive ($M_{\ast}\sim10^{11}\,\msun$), red, unobscured ellipticals with old
($\gtrsim$1\,Gyr) stellar populations \citep{Rupke2024,Coil2024,Norris2022}. Our host is
at the lower-mass end of that distribution but otherwise indistinguishable, which
strengthens the case for a genuine association between the galaxy and the bubbles. By inspecting the radio image, we found a weak ($\sim0.5$ mJy), spot-like radio source that is spatially coincident with the position of the host galaxy (see Fig. \ref{fig:bubble}). This resembles some ORCs, which exhibit radio cores associated with their host galaxies \citep[e.g.][]{Koribalski2021,Norris2022}. The origin of the radio emission is not clear. Using the SFR derived from SED fitting, we estimate the rest-frame 1.4~GHz radio luminosity produced by star formation to be $L_{1.4\,\mathrm{GHz}}\sim4.0\times10^{19}\,\mathrm{W}\,\mathrm{Hz}^{-1}$ from the SFR--radio luminosity correlation \citep{kennicutt2012}. This is much smaller than the observed $L_{1.4\,\mathrm{GHz}}\sim1.2\times10^{21}\,\mathrm{W}\,\mathrm{Hz}^{-1}$ for the host galaxy, suggesting a possible AGN jet. Higher spatial resolution and sensitivity are required to test the jet scenario.

\begin{figure}[t]
    \centering
    \includegraphics[width=0.9\columnwidth]{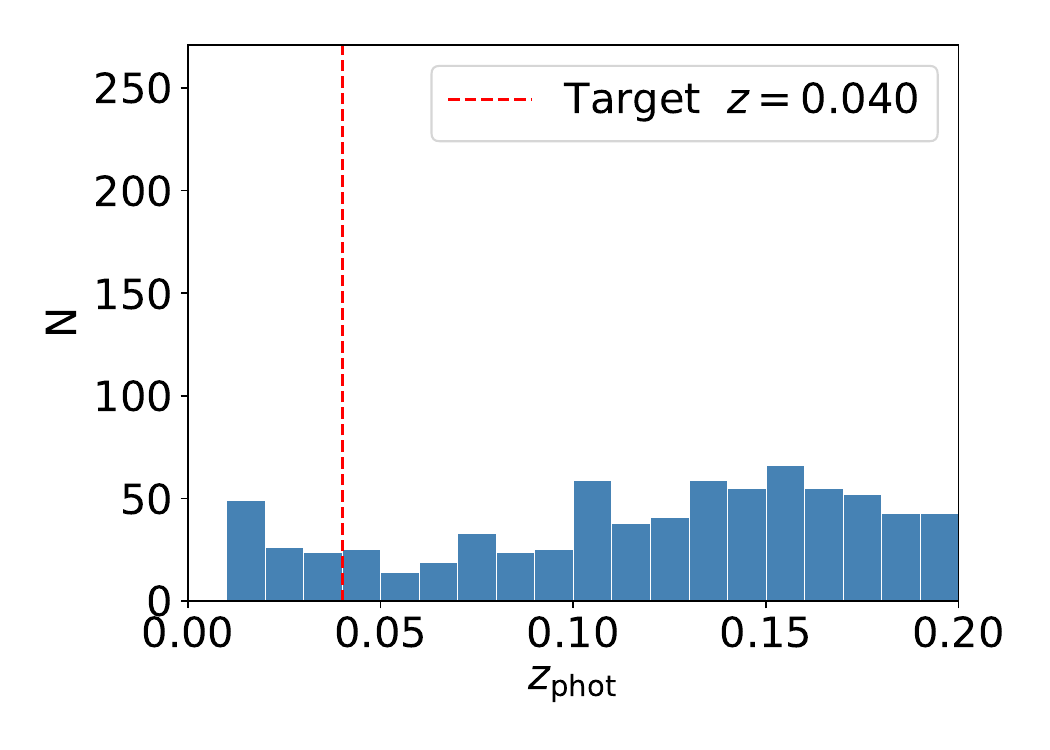}
    \caption{Photometric-redshift distribution of galaxies within 30\,arcmin of
    LEDA~217397, from the catalogue of \citet{Tian2026}. The red dashed line marks the
    host redshift, $z=0.040$; no overdensity is seen at the host redshift. Only redshift range of 0--0.2 is shown.}
    \label{fig:zhist}
\end{figure}

\begin{figure}[t]
    \centering
    \includegraphics[width=0.9\columnwidth]{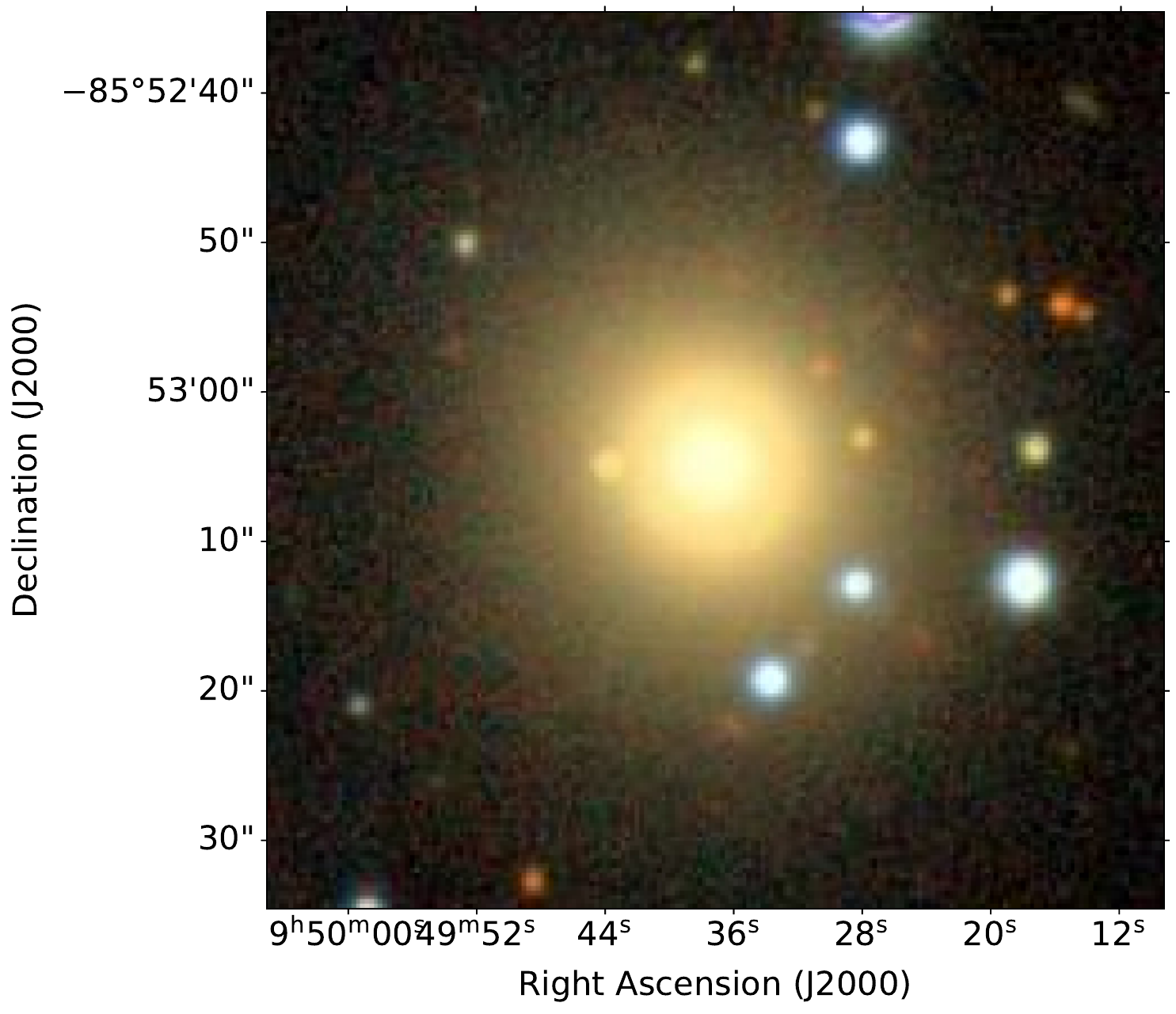}
    \caption{A close view with a size of $60^{\prime\prime}\times60^{\prime\prime}$ of the host galaxy.}
    \label{fig:zoom_host}
\end{figure}

To assess the large-scale environment we examined the photometric redshifts of all
galaxies within 30\,arcmin ($\simeq$1.4\,Mpc) of the host, drawn from the neural-network
photo-$z$ catalogue of \citet{Tian2026}. The redshift histogram in Fig.~\ref{fig:zhist} shows no excess of galaxies at $z=0.040$; the counts per bin
near the host redshift ($\sim$25--50 per $\Delta z=0.01$ bin) are comparable to the
adjacent field population, and no significant spike is present at the target redshift.
LEDA~217397 therefore does not reside in a rich group or cluster, although we caution
that photo-$z$ uncertainties ($\sigma_{z}\sim0.02$--0.05) would smear the signature of a
poor group over several bins. This contrasts with the statistical tendency of the
original ORC hosts to lie in dense environments \citep{Norris2021b}. However, a careful inspection of the DESI optical image of the host can identify some very close companions, as shown in Fig.~\ref{fig:zoom_host}. It is unclear whether they are physically associated with LEDA~217397.

\subsection{A starburst-driven wind origin?}\label{sec:wind}

Starburst-driven winds are a natural candidate for kpc-to-100-kpc scale bubbles: a
concentrated episode of star formation drives a fast ($10^{3}$\,km\,s$^{-1}$) hot wind
that shocks the CGM, sweeping up a slow, dense shell -- a geometry known to produce
limb-brightened, approximately spherical shells \citep{Weaver1977,Lochhaas2018}. A related ``explosive outflow remnant'' picture has been applied to ORCs by \citet{Fujita2024}; we note, however, its $\sim$$10^{60}$\,erg energy budget is more naturally supplied by an AGN than by a starburst, as those authors acknowledge, so we defer further discussion to Sect.~\ref{sec:jet}.

We now derive the energy that any successful model must supply, since it is this number
that most strongly constrains the wind scenario. For the
emitting and cavity volume $V$ we adopt the union of
the two fitted shells of Sect.~\ref{sec:properties}: two spheres of radii
$R_{\rm SW}=R_{\rm NE}=114$\,kpc whose centres are separated by $d\simeq171$\,kpc. Neglecting the small overlap between the two
spheres, we take the volume to be simply the sum of the two sphere volumes,
\begin{equation}\label{eq:volume}
V \;=\; \frac{4\pi}{3}\Bigl(R_{\rm SW}^{3}+R_{\rm NE}^{3}\Bigr)
\;\simeq\; 1.2\times10^{7}\ {\rm kpc^{3}} \\ \;\simeq\; 3.6\times10^{71}\ {\rm cm^{3}}.
\end{equation}

If the relativistic plasma actually fills only a fraction $\phi$ of this volume (for
example a pair of thin shells), the equipartition energies derived below
scale weakly with that fraction ($E_{\min}\propto\phi^{1/2}$); we adopt $\phi=1$, consistent with the diffuse emission
detected throughout the interior of both shells.

\emph{(i) Minimum (equipartition) energy of the synchrotron-emitting plasma.} The radio
emission is produced by relativistic electrons gyrating in a magnetic field of strength
$B$. For a given observed luminosity, the energy stored in electrons, $E_{\rm e}$, is a
steeply decreasing function of $B$ (a weaker field needs more electrons to produce the
same emission), while the magnetic energy $E_{\rm B}=U_{\rm B}V=(B^{2}/8\pi)V$, with
$U_{\rm B}$ the magnetic energy density, increases with $B$; the total
$E_{\rm e}+E_{\rm B}$ therefore has a minimum, which occurs close to equipartition
between the two \citep{Beck2005}. To evaluate it, we model the electron population as a
power law in Lorentz factor $\gamma$,
\begin{equation*}
N(\gamma)=N_{0}\,\gamma^{-p},
\end{equation*}
where $N(\gamma)\,{\rm d}\gamma$ is the number of electrons per unit volume with Lorentz
factor between $\gamma$ and $\gamma+{\rm d}\gamma$, $N_{0}$ is a normalization constant,
and $p$ is the electron energy index. The index follows directly from the observed
spectrum. For optically thin synchrotron emission, $S_{\nu}\propto\nu^{-(p-1)/2}$, so
our measured $\alpha=-1.04$ implies $p=2|\alpha|+1\simeq3.1$. The synchrotron power of
one electron is $P_{\rm syn}(\gamma)=(4/3)\sigma_{\rm T}c\,U_{\rm B}\gamma^{2}$, where
$\sigma_{\rm T}$ is the Thomson cross-section and $c$ the speed of light. Integrating
over the whole electron population, the total synchrotron luminosity is
\begin{equation}\label{eq:lsyn}
L_{\rm syn} \;=\; \frac{4}{3}\,\sigma_{\rm T}\,c\,U_{\rm B}\!
\int_{\gamma_{1}}^{\gamma_{2}}\! N(\gamma)\,\gamma^{2}\,{\rm d}\gamma \;\simeq\;
1.5\times10^{40}\ {\rm erg\,s^{-1}},
\end{equation}
where the numerical value is obtained by integrating the observed power-law spectrum from
10\,MHz to 10\,GHz. Taking $\gamma_{1}=10$ and $\gamma_{2}=10^{5}$ as the adopted lower and upper cut-off
Lorentz factors of the electron spectrum and setting the electron energy density equal to the magnetic one,
$U_{\rm e}=U_{\rm B}$, with
$U_{\rm e}=\int_{\gamma_{1}}^{\gamma_{2}}N(\gamma)\,\gamma m_{\rm e}c^{2}\,{\rm d}\gamma$
($m_{\rm e}c^{2}$ being the electron rest-mass energy), and solving Eq.~\eqref{eq:lsyn}
within the two-shell volume of Eq.~\eqref{eq:volume} gives
\begin{equation}\label{eq:emin}
B_{\rm eq}\simeq2.0\,\mu{\rm G},\qquad
E_{\min}\simeq1.1\times10^{59}\ {\rm erg},
\end{equation}
where $B_{\rm eq}$ is the equipartition magnetic field and $E_{\min}$ the corresponding
minimum total (electrons plus magnetic field) energy. If relativistic protons carry $K$
times more energy than the electrons (as in Galactic cosmic rays, where $K\sim100$),
minimizing the total (particle plus magnetic) energy gives a field
$B_{\rm eq}\propto(1+K)^{1/4}$ and a minimum total energy $E_{\min}\propto(1+K)^{1/2}$ \citep{Beck2005},
i.e.\ $B_{\rm eq}\simeq6.1\,\mu$G and $E_{\min}\simeq1.1\times10^{60}$\,erg for $K=100$;
$1.1\times10^{59}$\,erg is thus a firm lower limit.

\emph{(ii) Cooling-time.} The electrons radiating at 944\,MHz have Lorentz
factors $\gamma\simeq2\times10^{4}$ in a $\mu$G-strength field (adopting the convention
that the observed frequency is $\simeq$$0.29$ times the critical frequency; the
$\nu\simeq\nu_{\rm c}$ convention would give $\gamma\simeq1\times10^{4}$), and lose
energy to both
synchrotron radiation and inverse-Compton (IC) scattering of cosmic-microwave-background
(CMB) photons, whose energy density is equivalent to that of a magnetic field of
strength $B_{\rm CMB}=3.24(1+z)^{2}\simeq3.5\,\mu$G. The radiative lifetime,
\begin{equation}\label{eq:tcool}
t_{\rm cool}\;\simeq\;1.1\times10^{9}\,
\frac{B^{1/2}}{B^{2}+B_{\rm CMB}^{2}}\,\nu_{\rm GHz}^{-1/2}\ {\rm yr}
\qquad(B\ {\rm in}\ \mu{\rm G}),
\end{equation}
where $\nu_{\rm GHz}$ is the observing frequency in GHz, is only
$\sim$$10^{8}$\,yr at 944\,MHz over the whole plausible range $B\simeq2$--7\,$\mu$G
(IC losses on the CMB in fact dominate for $B\lesssim3.5\,\mu$G). Whatever accelerated
the electrons must therefore have done so no more than $\sim$$10^{8}$\,yr ago, or must
still be active.

The starburst-wind scenario can now be tested against these numbers. The mechanical
(kinetic) power of a starburst-driven wind is calibrated by stellar-population synthesis
models, which add up the energy injected by stellar winds and supernovae. For continuous
star formation with a standard initial mass function, \textsc{starburst99} models
\citep{Leitherer1999} show that, once the starburst has reached steady state, the mechanical luminosity saturates at
\begin{equation}\label{eq:wind}
L_{\rm w}\;\simeq\;7\times10^{41}\,
\Bigl(\frac{\sfr}{\msun\,{\rm yr}^{-1}}\Bigr)\ {\rm erg\,s^{-1}}
\end{equation}
\citep[see also the review by][]{Veilleux2005}. At the host's current
$\sfr\simeq0.025\,\msun$\,yr$^{-1}$, Eq.~\eqref{eq:wind} delivers
$\sim$$1.7\times10^{40}$\,erg\,s$^{-1}$, and accumulating the required $10^{59}$\,erg would
take an unreasonably $\sim$220\,Gyr. This far exceeds
the $\simeq$1.3\,Gyr expansion time of the shells (a distance of $\sim$200\,kpc, half of the size of the radio structure, travelled at a plausible CGM sound speed of $\sim$150\,km\,s$^{-1}$). More severely, the $\sim$$10^{8}$\,yr
radiative lifetime of the emitting electrons derived above means that wind earliest-accelerated electrons would have long cooled beyond detectability at 944\,MHz given the $\gtrsim$220\,Gyr required to inflate the bubbles. A short, intense ancient burst forming a significant fraction of the host's
$9\times10^{10}\,\msun$ of stars could in principle inject several $10^{59}$\,erg, but the
\textsc{cigale} star-formation history shows no such burst within the last several Gyr, Fig.~\ref{fig:sfh}.
The two-circle morphology provides a further argument against a wind. A single starburst
episode drives a single, host-centred bubble, whereas we observe two overlapping shells.
We therefore exclude the starburst-wind origin.

\subsection{A merger-driven shock origin?}\label{sec:merger}

The most actively developed ORC model in recent years invokes shocks driven into the CGM
by galaxy or halo mergers. Cosmological simulations show that mergers in
$10^{12}$--$10^{13}\,\msun$ halos naturally produce internal shocks with the sizes,
circular morphologies and Mach numbers ($\mathcal{M}\sim2$--3) required to match the ORC
population, with the host settling as a quenched early-type galaxy by the time the shell
reaches ORC dimensions \citep{Dolag2023,Ivleva2026}. Observationally, the Cloverleaf ORC
system traces a galaxy-group merger in progress \citep{Bulbul2024}; the Physalis and double-shell systems show pairs of shells consistent with merger
shocks in the CGM of interacting early-type galaxies
\citep{Koribalski2024b,Koribalski2026a}. A related suggestion is a single cataclysmic
blast wave from the coalescence of two SMBHs in the host nucleus
\mbox{\citep{Koribalski2021,Norris2022}}, possibly aided by the accumulated energy of tidal
disruption events \citep{Omar2022b}; such events release $10^{55}$--$10^{59}$\,erg, in
the range required by our energetics estimate.

Several quantitative aspects of our bubbles fit this framework. The integrated spectral
index $\alpha=-1.04$, interpreted as the injection index of diffusive shock acceleration,
corresponds to a moderate shock Mach number ($\mathcal{M}\simeq2$), and even if the
observed spectrum is instead a cooled segment above a cooling break (injection index
$\alpha_{\rm inj}\simeq-0.6$), a strong shock with $\mathcal{M}\approx4.5$--5 would be implied, both within the range of merger shocks found in cosmological simulations \citep[][]{Dolag2023,Ivleva2026}. The host is an elliptical, itself a probable merger remnant, with
a stellar mass and quiescent state matching the simulated hosts at the time their shells
are most prominent \citep{Dolag2023}. And the energy budget, $\sim$10$^{59}$\,erg
(Sect.~\ref{sec:wind}), is well within what a galaxy-scale merger releases in the CGM.

It is important to ask whether a merger shock can really account for the \emph{morphology} of our source, which is unusually regular. The emission
decomposes into two nearly perfect circular shells of similar radius. A close inspection of the published simulations and
merger-interpreted systems suggests this is not what mergers generically produce. In the
cosmological simulations of \citet{Dolag2023} and \citet{Ivleva2026}, merger-driven
shocks are typically aspherical, propagate along preferred (filamentary) directions, and
appear as open arcs or partial rings that close into a circle only under favourable
projection; the simulated ORC analogues are consequently irregular
rather than geometrically round. The observed merger systems likewise show irregular or
offset structures. For instance, the Cloverleaf ORC is a distorted leaf-shaped shell
\citep{Bulbul2024}. A single shock propagating through
a strongly inhomogeneous CGM could in principle break into two shell-like segments, but producing two shells that are individually this round, of comparable size, arranged symmetrically about the host, appears contrived in that picture. We conclude that a merger shock is unlikely to be the agent that \emph{shaped} the two shells, although it
cannot be excluded as a contributor -- for example by re-energising pre-existing
relativistic plasma in an already-formed bubble \citep[the ``phoenix'' idea
of][]{Shabala2024}.

X-ray emission is expected from shock--gas interaction. In the simulations of \citet{Dolag2023}, merger-driven shocks show up as steep surface-brightness jumps in the thermal X-ray emission of the circumgalactic gas (their Fig.~10). However, the predicted surface brightness is below the reach of present-day X-ray instruments by more than an order of magnitude --- with additional large uncertainties from the CGM metallicity, resonant scattering and galaxy-formation physics --- so that detecting such emission would require next-generation instruments such as the Line Emission Mapper \citep[LEM;][]{Dolag2023}. Our X-ray view of the bubbles is fully consistent with this expectation. We cross-matched the bubbles with the source catalogues of the eROSITA all-sky surveys \citep[3eRASS;][]{Predehl2021,Ramos2026}. No diffuse X-ray emission is detected from the bubbles or their shells: at the position of the bubbles, the $3\sigma$ upper limit on the 0.2--2.3\,keV flux is $3.7\times10^{-14}$\,erg\,s$^{-1}$\,cm$^{-2}$ (after correcting for Galactic absorption), corresponding to a luminosity limit of $1.4\times10^{41}$\,erg\,s$^{-1}$ at $z=0.040$. Only two faint, low-significance point sources are catalogued within the bubbles boundary: X1 ( 3eRASS~J094929.5$-$855310 with \texttt{det\_like\_0} = 6.1) and X2 (3eRASS~J094915.2$-$855547 with \texttt{det\_like\_0} = 9.7), whose positions and $\sim$$5^{\prime\prime}$ positional uncertainties are marked in Fig.~\ref{fig:bubble}(c). After correcting for Galactic absorption, their 0.2--2.3\,keV fluxes are $(1.9\pm0.8)\times10^{-14}$\,erg\,s$^{-1}$\,cm$^{-2}$ for X1, and $(2.1\pm0.8)\times10^{-14}$\,erg\,s$^{-1}$\,cm$^{-2}$ for X2; if the sources lie at the redshift of the bubbles, these correspond to 0.2--2.3\,keV luminosities of $(7.1\pm3.0)\times10^{40}$ and $(7.8\pm3.0)\times10^{40}$\,erg\,s$^{-1}$, respectively. Both sources are point-like at the eROSITA resolution, so they cannot represent the extended, shock-heated CGM emission predicted by the merger-shock model; given their low detection significance, they are most plausibly background AGNs, X-ray binaries, or even statistical fluctuations, and any physical association with the bubbles remains unclear. The absence of diffuse shock emission therefore neither supports nor excludes a merger-shock origin: a shock of the kind simulated by \citet{Dolag2023} would not have been detectable with eROSITA.

One additional tension is environmental. Fig.~\ref{fig:zhist} shows no group-scale
overdensity around LEDA~217397 (although a few very close companions of uncertain association are visible in Fig.~\ref{fig:zoom_host}), whereas the best-studied merger-shock ORCs reside in
groups or clusters \citep{Bulbul2024,Koribalski2024a,Koribalski2024b}. However, the tension is
not fatal. The
merger may have completed long enough ago that companions have coalesced or fallen below
the photometric limits; and the photo-$z$ smearing noted in Sect.~\ref{sec:host} could
hide a poor group. Spectroscopy of a number of galaxies with
$z_{\rm phot}\simeq0.03$--0.05 within 30\,arcmin would settle whether a small group is
present. We also note that the elliptical morphology of the host itself points to a past
major merger, which could have powered the bubbles through the coalescence of the two
SMBHs \citep{Koribalski2021,Norris2022}. Here, however, a spherical blast wave from SMBH coalescence would
produce a \emph{single} host-centred shell, whereas we have shown
(Sect.~\ref{sec:properties}) that the emission is not a single ring. A purely isotropic nuclear explosion is therefore ruled
out as the shaping agent, just as the halo-scale shock was. The way out is that SMBH
coalescence need not be isotropic. General-relativistic simulations of magnetized
mergers show that the remnant promptly launches a pair of oppositely directed,
transient jets \citep[via the Blandford--Znajek mechanism;][]{Blandford1977,
Palenzuela2010}. Such a bipolar nuclear outburst would inflate two lobes on opposite
sides of the host, precisely the double shell geometry we observe, and
in this situation the SMBH-merger scenario transfers into the relic-AGN jet
scenario discussed next. If the ``merger'' channel is correct, it is thus the
\emph{bipolar} nuclear outburst, not the spherical blast wave and not the CGM shock,
that we are seeing.

\subsection{An AGN jet-inflated bubble origin?}\label{sec:jet}

The third family of models links ORCs to past AGN jet activity. \citet{Lin2024} showed
with cosmic-ray MHD simulations that powerful, long-duration jets dominated by cosmic-ray
protons inflate oblate bubbles that, seen end-on, reproduce the $\sim$300--600\,kpc
sizes and edge-brightened morphology of ORCs (the limb brightening arising naturally from
hadronic collisions at the bubble--CGM interface). \citet{Shabala2024} instead proposed
that ORCs are remnant lobes of powerful radio galaxies re-energised by a passing shock, which produces both filled and edge-brightened
morphologies; and end-on projection of classical double radio sources or precessing jets
can yield ring-like structures in special geometries
\citep{Koribalski2021,Nolting2023}.

Our source offers several hooks for this scenario. The steep, uniform spectrum
($\alpha=-1.04$) is exactly what is expected for aged lobe plasma, and the two
approximately circular components, both of $\simeq$228\,kpc diameter, could be
the two lobes of a jet.

The required jet energy, of order the bubbles
energy $\sim$10$^{59}$\,erg (Sect.~\ref{sec:wind}), corresponds to a time-averaged jet
power of only
$\sim$$10^{43}$\,erg\,s$^{-1}$ sustained over $\sim$$3\times10^{8}$\,yr, entirely
ordinary for a massive elliptical. We repeat that a weak, spot-like radio source that is spatially coincident with the position of the host galaxy was found (see Fig. \ref{fig:bubble}), indicating the existence of a possible AGN jet. The absence of any AGN signature in the WISE colours or the SED, argue that the AGN is currently off, consistent with that the visible shells are fossils or relic-lobes. We stress that the morphological
argument developed in Sect.~\ref{sec:merger} also works in favour of this family of
models in which jet-inflated bubbles and radio lobes are inflated from the nucleus outwards and
are therefore naturally round, host-centred and -- for a bipolar source -- double, which
is precisely the geometry we observe.

We note that the merger and AGN scenarios may not be independent. If the past
major merger that made the host an elliptical also drove a phase of strong nuclear
accretion (or launched merger jets; \citealt{Palenzuela2010}), then the bubbles could be
the fossil of the resulting bipolar outburst, regardless of whether the ultimate energy
reservoir is labelled ``SMBH merger'' or ``AGN''. Distinguishing a relic-AGN bubble
from a merger-powered outburst will require (i) higher-resolution
radio imaging to search for a compact core, fossil jets or lobe substructure; (ii)
broad-band observations and polarimetry, ordered tangential fields in the rim
would favour shock compression \citep{Norris2022}, while double-ring polarised
structure could betray a lobe origin \citep{Taziaux2025}; and (iii) X-ray observations
to detect a cavity, a shock-heated shell, or the hot CGM into which a shock must be
propagating \citep{Bulbul2024,Koribalski2024b}. With its low redshift, high flux density
and large angular size, the bubbles are arguably the best laboratory yet available for
these tests.

\section{Summary}\label{sec:summary}

We have reported the discovery of a pair of large, low-surface-brightness radio bubbles in ASKAP
EMU 944\,MHz continuum data, centred on the elliptical galaxy LEDA~217397 at
$z=0.040$. Our main findings are as follows.

\begin{enumerate}[leftmargin=1.4em,itemsep=2pt]
\item The radio structure spans $\sim$8.4\,arcmin ($\sim$399\,kpc) and is composed of two partly
overlapping, approximately circular shells both with radii of $\sim$114\,kpc.
The host galaxy sits almost at the geometrical centre of the whole structure.

\item The integrated flux density is $56.8\pm2.9$\,mJy at 944\,MHz, corresponding to $L_{944}\simeq2.1\times10^{23}$\,W\,Hz$^{-1}$ and $L_{1.4}\simeq1.4\times10^{23}$\,
W\,Hz$^{-1}$. In size, surface brightness, luminosity and edge-brightened morphology the
source closely resembles the ORC population, while being one of the nearest and the
brightest members to date (Sect.~\ref{sec:properties}).

\item Combining EMU with GLEAM-X data at 88--185\,MHz, the integrated spectrum is a
steep power law with $\alpha=-1.04\pm0.04$, comparable to the spectra of confirmed ORCs.
The spatially resolved spectral indices indicate an exterior flattening but with large uncertainties; deeper
observations are needed to explore the spectral structure (Sect.~\ref{sec:spix}).

\item The host is a massive ($\log M_{\ast}/\msun=10.97\pm0.09$), quiescent
($\sfr=0.025\pm0.083\,\msun$\,yr$^{-1}$) elliptical on the red sequence, with WISE colours
incompatible with a current AGN, an old stellar population, and no group-scale
overdensity within 30\,arcmin. These properties closely match those of confirmed ORC
hosts (Sect.~\ref{sec:host}).

\item Of the three origin scenarios examined, a recent starburst-driven wind is
disfavoured: the host's current and recent star formation is far too weak to supply the
required $\gtrsim$$10^{59}$\,erg and a single wind
would not produce the two-circle morphology. A halo-scale merger shock can match the
energy and spectral index, but struggles to produce two nearly perfect, host-centred
circular shells, and no group-scale overdensity is present; a spherical
supermassive-black-hole merger blast wave is likewise excluded, since it would produce a
single shell rather than the observed two. The most natural explanation is
therefore a bipolar nuclear outburst -- a relic AGN jet episode, possibly triggered by a
supermassive-black-hole merger \citep{Palenzuela2010} -- with a later shock possibly
re-energising the ejected plasma (Sect.~\ref{sec:jet};
\citealt{Shabala2024,Koribalski2024b}).
\end{enumerate}

Whatever its precise origin, this object demonstrates that ASKAP's wide-field continuum surveys
are a rich source of serendipitous discoveries, and that large radio shells
around nearby galaxies may be more common than the classical, more distant ORC sample
suggests. The decisive next steps are deep, broad-band, higher-resolution radio
observations ( such as very long baseline interferometry observations to search for a core, jets and spectral gradients associated with the host galaxy and the galaxy ``A''), polarimetry (to map thes
magnetic-field geometry), spectroscopy of the host and neighbouring galaxies (to confirm
the redshift, probe the host's gas content and search for a hidden group), and X-ray
follow-up (to detect the hot gas associated with any shock or cavity). Each of these is
uniquely feasible for a source this bright and this nearby.

%% Also note that the akcnowlodgment environment does not support long amounts of text. If you have a lot of people and institutions to acknowledge, do not use this command. Instead, create a new \section{Acknowledgments}.
\section*{Acknowledgments}

We thank the anonymous referee for the comments that have improved the quality of this paper and Kathryn Ross for sharing us the GLEAM-X data.

This work is supported by the National Science and Technology Major Project of China (No. 2024ZD1100601), the National SKA Program of China (No. 2025SKA0130100). RZS acknowledges the support from the Shanghai Super Postdoctoral Incentive Program (No. 2025322), and the China Postdoctoral Science Foundation (Grant No. 2025M783232). MFG is supported by the National Science Foundation of China (grant 12473019), the National SKA Program of China (Grant No. 2022SKA0120102)，the Shanghai Pilot Program for Basic Research-Chinese Academy of Science, Shanghai Branch (JCYJ-SHFY-2021-013), and the China Manned Space Project with No. CMS-CSST-2025-A07.  L.C. acknowledges support by the National SKA Program of China (Grant No.2022SKA0120102), the National Key R\&D program of China under the grant 2024YFA1611403, 2024YFA1611401, 2024YFA1611402, 2024YFA1611404 and Shanghai Pilot Program for Basic Research, Chinese Academy of Science, Shanghai Branch (JCYJ-SHFY-2021-013). G.M. is supported by the National Science Foundation of China (No. 12473013). F.G. acknowledges the support from the Excellent Youth Team Project of the Chinese Academy of Sciences (No. YSBR-061), the National Natural Science Foundation of China (No. 12473010). X.Z. acknowledges the support from the China Postdoctoral Science Foundation (General Program, Grant No. 2024M753368), the National Natural Science Foundation of China (Young Scientists Fund, Class C, Grant No. 12503021), the Shanghai Sailing Program (Grant No. 24YF2754400), the Shanghai Super Postdoctoral Incentive Program, and the Special Research Assistant Project of the Chinese Academy of Sciences. WWZ is supported by the Shanghai Natural Science Foundation Youth Project ( Grant No .25ZR1402546), Strategic Priority Research Program of t he Chinese Academy of Sciences ( Grant No . XDB 0800302) and the National Key Research and Development Program of China ( Grant No .2025 Y FA1614102).

The Australian SKA Pathfinder is part of the Australia Telescope
National Facility (https://ror.org/05qajvd42), which is managed by CSIRO. Operation of ASKAP is funded by the Australian Government with support from the National Collaborative Research Infrastructure Strategy. ASKAP uses the resources of the Pawsey Supercomputing Centre. Establishment of ASKAP, the Murchison Radio-astronomy Observatory, and the Pawsey Supercomputing Centre are initiatives of the Australian Government, with support from the Government of Western Australia and the Science and Industry Endowment Fund. We acknowledge the Wajarri Yamatji people as the traditional owners
of the Observatory site.

This work uses data obtained from Inyarrimanha Ilgari
Bundara/the Murchison Radio-astronomy Observatory. We
acknowledge the Wajarri Yamaji People as the Traditional
Owners and native title holders of the Observatory site.
Establishment of CSIRO's Murchison Radio-astronomy
Observatory and the Pawsey Supercomputing Center are
initiatives of the Australian Government, with support from
the Government of Western Australia and the Science and
Industry Endowment Fund. Support for the operation of the
MWA is provided by the Australian Government (NCRIS),
under a contract to Curtin University administered by
Astronomy Australia Limited.

This paper is based on data acquired at the ANU 2.3-metre telescope. The automation of the telescope was made possible through an initial grant provided by the Centre of Gravitational Astrophysics and the Research School of Astronomy and Astrophysics at the Australian National University and through a grant provided by the Australian Research Council through LE230100063. We acknowledge the traditional custodians of the land on which the telescope stands, the Gamilaraay people, and pay our respects to elders past and present.

The Legacy Surveys consist of three individual and complementary projects: the Dark Energy Camera Legacy Survey (DECaLS; Proposal ID \#2014B-0404; PIs: David Schlegel and Arjun Dey), the Beijing-Arizona Sky Survey (BASS; NOAO Prop. ID \#2015A-0801; PIs: Zhou Xu and Xiaohui Fan), and the Mayall z-band Legacy Survey (MzLS; Prop. ID \#2016A-0453; PI: Arjun Dey). DECaLS, BASS and MzLS together include data obtained, respectively, at the Blanco telescope, Cerro Tololo Inter-American Observatory, NSF's NOIRLab; the Bok telescope, Steward Observatory, University of Arizona; and the Mayall telescope, Kitt Peak National Observatory, NOIRLab. Pipeline processing and analyses of the data were supported by NOIRLab and the Lawrence Berkeley National Laboratory (LBNL). The Legacy Surveys project is honored to be permitted to conduct astronomical research on Iolkam Du'ag (Kitt Peak), a mountain with particular significance to the Tohono O'odham Nation.

NOIRLab is operated by the Association of Universities for Research in Astronomy (AURA) under a cooperative agreement with the National Science Foundation. LBNL is managed by the Regents of the University of California under contract to the U.S. Department of Energy.

This project used data obtained with the Dark Energy Camera (DECam), which was constructed by the Dark Energy Survey (DES) collaboration. Funding for the DES Projects has been provided by the U.S. Department of Energy, the U.S. National Science Foundation, the Ministry of Science and Education of Spain, the Science and Technology Facilities Council of the United Kingdom, the Higher Education Funding Council for England, the National Center for Supercomputing Applications at the University of Illinois at Urbana-Champaign, the Kavli Institute of Cosmological Physics at the University of Chicago, Center for Cosmology and Astro-Particle Physics at the Ohio State University, the Mitchell Institute for Fundamental Physics and Astronomy at Texas A\&M University, Financiadora de Estudos e Projetos, Fundacao Carlos Chagas Filho de Amparo, Financiadora de Estudos e Projetos, Fundacao Carlos Chagas Filho de Amparo a Pesquisa do Estado do Rio de Janeiro, Conselho Nacional de Desenvolvimento Cientifico e Tecnologico and the Ministerio da Ciencia, Tecnologia e Inovacao, the Deutsche Forschungsgemeinschaft and the Collaborating Institutions in the Dark Energy Survey. The Collaborating Institutions are Argonne National Laboratory, the University of California at Santa Cruz, the University of Cambridge, Centro de Investigaciones Energeticas, Medioambientales y Tecnologicas-Madrid, the University of Chicago, University College London, the DES-Brazil Consortium, the University of Edinburgh, the Eidgenossische Technische Hochschule (ETH) Zurich, Fermi National Accelerator Laboratory, the University of Illinois at Urbana-Champaign, the Institut de Ciencies de l’Espai (IEEC/CSIC), the Institut de Fisica d’Altes Energies, Lawrence Berkeley National Laboratory, the Ludwig Maximilians Universitat Munchen and the associated Excellence Cluster Universe, the University of Michigan, NSF’s NOIRLab, the University of Nottingham, the Ohio State University, the University of Pennsylvania, the University of Portsmouth, SLAC National Accelerator Laboratory, Stanford University, the University of Sussex, and Texas A\&M University.

BASS is a key project of the Telescope Access Program (TAP), which has been funded by the National Astronomical Observatories of China, the Chinese Academy of Sciences (the Strategic Priority Research Program “The Emergence of Cosmological Structures” Grant \# XDB09000000), and the Special Fund for Astronomy from the Ministry of Finance. The BASS is also supported by the External Cooperation Program of Chinese Academy of Sciences (Grant \# 114A11KYSB20160057), and Chinese National Natural Science Foundation (Grant \# 12120101003, \# 11433005).

The Legacy Survey team makes use of data products from the Near-Earth Object Wide-field Infrared Survey Explorer (NEOWISE), which is a project of the Jet Propulsion Laboratory/California Institute of Technology. NEOWISE is funded by the National Aeronautics and Space Administration.

The Legacy Surveys imaging of the DESI footprint is supported by the Director, Office of Science, Office of High Energy Physics of the U.S. Department of Energy under Contract No. DE-AC02-05CH1123, by the National Energy Research Scientific Computing Center, a DOE Office of Science User Facility under the same contract; and by the U.S. National Science Foundation, Division of Astronomical Sciences under Contract No. AST-0950945 to NOAO.

This work is based on data from eROSITA, the soft X-ray instrument aboard SRG, a joint Russian-German science mission supported by the Russian Space Agency (Roskosmos), in the interests of the Russian Academy of Sciences represented by its Space Research Institute (IKI), and the Deutsches Zentrum f\"{u}r Luft- und Raumfahrt (DLR). The SRG spacecraft was built by Lavochkin Association (NPOL) and its subcontractors, and is operated by NPOL with support from the Max Planck Institute for Extraterrestrial Physics (MPE). The development and construction of the eROSITA X-ray instrument was led by MPE, with contributions from the Dr. Karl Remeis Observatory Bamberg \& ECAP (FAU Erlangen-Nuernberg), the University of Hamburg Observatory, the Leibniz Institute for Astrophysics Potsdam (AIP), and the Institute for Astronomy and Astrophysics of the University of T\"{u}bingen, with the support of DLR and the Max Planck Society. The Argelander Institute for Astronomy of the University of Bonn and the Ludwig Maximilians Universität Munich also participated in the science preparation for eROSITA.

\

%% To help institutions






\bibliography{orc}{}
\bibliographystyle{aasjournal}


\end{CJK*}
\end{document}